\PassOptionsToPackage{hyperfootnotes=false}{hyperref} 
\documentclass[fleqn,usenatbib]{mnras}

\usepackage{newtxtext,newtxmath}
\SetSymbolFont{bmisymbols}{normal}{OML}{ntxmi}{b}{it} 

\usepackage[T1]{fontenc}

\DeclareRobustCommand{\VAN}[3]{#2}
\let\VANthebibliography\thebibliography
\def\thebibliography{\DeclareRobustCommand{\VAN}[3]{##3}\VANthebibliography}

\usepackage{graphicx}
\usepackage{amsmath}
\usepackage{orcidlink}
\usepackage{siunitx}
\usepackage[version=4]{mhchem}
\usepackage{multirow}

\newcounter{reaction}
\renewcommand{\thereaction}{R\arabic{reaction}}
\newcommand{\reactionprefix}{}
\renewcommand{\thereaction}{\reactionprefix\arabic{reaction}}
\newcommand{\reactiontag}{%
  \refstepcounter{reaction}(\thereaction)%
}

\newcounter{pathway}
\renewcommand{\thepathway}{P\arabic{pathway}}

\title[\ce{CH4}-to-\ce{CO} pathways and simplified schemes]{Spatial variability of \ce{CH4}-to-\ce{CO} conversion pathways in irradiated gas giant atmospheres: implications for chemical relaxation and 1D quench schemes}

\author[Zamyatina et al.]{Maria Zamyatina$^{{\orcidlink{0000-0002-9705-0535}},1}$\thanks{E-mail: m.zamyatina@exeter.ac.uk},
Éric H\'ebrard$^{{\orcidlink{0000-0003-0770-7271}},1}$,
Harry Baskett$^{{\orcidlink{0009-0003-1097-5792}},1}$,
Roméo Veillet$^{{\orcidlink{0009-0007-2485-0308}},1}$,
Duncan A. Christie$^{{\orcidlink{0000-0002-4997-0847}},2}$,
\newauthor
Nathan J. Mayne$^{{\orcidlink{0000-0001-6707-4563}},1}$,
Denis E. Sergeev$^{{\orcidlink{0000-0001-8832-5288}},3}$,
and Krisztian Kohary$^{{\orcidlink{0000-0003-0165-4885}},1}$
\\
$^{1}$Department of Physics and Astronomy, Faculty of Environment, Science and Economy, University of Exeter, Exeter EX4 4QL, UK\\
$^{2}$Max-Planck-Institut f\"{u}r Astronomie, K\"{o}nigstuhl 17, 69117 Heidelberg, Germany\\
$^{3}$School of Physics, University of Bristol, HH Wills Physics Laboratory, Tyndall Avenue, Bristol BS8 1TL, UK
}

\date{Accepted XXX. Received YYY; in original form ZZZ}

\pubyear{2026}

\begin{document}
\label{firstpage}
\pagerange{\pageref{firstpage}--\pageref{lastpage}}
\maketitle

\begin{abstract}

\noindent
Close-in gas giant atmospheres are in chemical disequilibrium.
Solving 3D kinetics to capture one source of this disequilibrium, transport-induced quenching, is computationally demanding, so simplifications are used.
General circulation models (GCMs) relax abundances on the timescale of a single conversion pathway.
1D models place a quench level and hold abundances vertically uniform above it.
We test both simplifications on \ce{CH4} against aerosol-free, solar-composition GCM simulations with thermochemical kinetics for HAT-P-11b, HD~189733b, HD~209458b and WASP-17b.
We find ten \ce{CH4}-to-\ce{CO} conversion pathways, varying with depth, longitude and planet.
A single pathway controls \ce{CH4} quenching on the three cooler planets; on WASP-17b two compete.
The \ce{CH4} quench level is 3D: where the atmosphere is hotter, or a faster pathway takes over, the conversion timescale shortens, the quench level rises and the quench-level abundance falls.
On WASP-17b the pathway switch lifts the quench level over the substellar tropics, deepening equatorial \ce{CH4} depletion.
A 1D quench treatment with the Zahnle \& Marley timescale typically locates the quench level within a factor of two in pressure for \ce{CH4} and about four for \ce{CO} on every planet.
It fails above the quench level, where every quench scheme holds abundances vertically uniform.
The minor carbon carrier, \ce{CH4} or \ce{CO} depending on the planet, is wrong by up to two orders of magnitude, on HD~189733b in opposite directions at different latitudes.
Quenching sets an abundance at the quench level but does not fix it thereafter; even without photochemistry, transport and conversion keep that abundance spatially structured.

\end{abstract}

\begin{keywords}
planets and satellites: atmospheres --
planets and satellites: composition --
planets and satellites: individual: HAT-P-11b, HD~189733b, HD~209458b, WASP-17b.
\end{keywords}

\section{Introduction}
\label{sec:intro}

Recent observations of close-in gas giants increasingly indicate that their observable atmospheres are shaped by disequilibrium chemistry \citep{Bell23_methane,Tsai23_photochemically,Beatty24_sulfur,Bell24_nightside,Dyrek24_so2,Yang24_simultaneous,Gressier25_jwsttst,Parmentier26_horizontal}.
This is expected because these planets experience a strong day–night irradiation contrast, which in addition to driving vigorous atmospheric circulation also promotes disequilibrium chemistry.
Among the affected species, \ce{CH4} is particularly diagnostic of disequilibrium.
In thermochemical equilibrium the relative abundances of \ce{CH4} and \ce{CO} vary strongly with pressure and temperature: carbon resides primarily in \ce{CH4} at lower temperatures and higher pressures, but shifts to \ce{CO} at higher temperatures and lower pressures.
Disequilibrium chemistry disrupts this partitioning.
Transport-induced quenching either enriches or depletes \ce{CH4}, depending on metallicity, \ce{C}/\ce{O} ratio, interior heat flux, and three-dimensional (3D) mixing \citep{Baeyens21_grid,Zamyatina24_quenchingdriven}.
Photochemistry destroys \ce{CH4} on the irradiated dayside \citep{Moses14_chemical,Tsai21_comparative,Baeyens22_grid}.
Together photochemistry and transport-induced quenching drive \ce{CH4} abundances away from equilibrium, and affect whether \ce{CH4} absorption contributes appreciably to planetary transmission and emission spectra.
Therefore, robust interpretation of \ce{CH4} signatures in planetary spectra requires general circulation models (GCMs) that account for disequilibrium chemistry.

Coupling large chemical kinetics networks to GCMs is computationally demanding, so simplified methods for treating disequilibrium chemistry were used.
\citet{Cooper06_dynamics} introduced the first such method for hot Jupiters.
In this method, \ce{CO} and \ce{CH4} were the only chemical species considered as being in disequilibrium, and for further simplification, only \ce{CO} was evolved as a tracer, with its abundance relaxed towards its equilibrium abundance on a timescale of \ce{CO}-to-\ce{CH4} conversion.
The timescale of this conversion was computed based on a series of reactions of \citet{Yung88_estimation} that described the reduction of \ce{CO} to \ce{CH4} proceeding as follows:
\renewcommand{\reactionprefix}{C}
\setcounter{reaction}{0}
\par\noindent
\begin{tabular*}{\columnwidth}{@{}r@{\;}l@{\extracolsep{\fill}}r@{}}
    \ce{CO + H2 + M} & \ce{-> H2CO + M} & \reactiontag \\
    \ce{H2CO + H + M} & \ce{-> CH3O + M} & \reactiontag\label{rxn:C2} \\
    \ce{CH3O + H2} & \ce{-> CH3OH + H} & \reactiontag \\
    \ce{CH3OH + H} & \ce{-> CH3 + H2O} & \reactiontag \\
    \ce{CH3 + H + M} & \ce{-> CH4 + M} & \reactiontag \\
\end{tabular*}
Because reaction (C2) breaks the \ce{C=O} bond, this reaction was identified as the rate-limiting step that controls the speed and hence the timescale of \ce{CO}-to-\ce{CH4} conversion.
The rate coefficient of (C2) has never been measured, so it was inferred at the time\footnote{More recent data are available from \citet{Hippler01_detailed} and \citet{Nguyen21_thermal}.} from the reverse rate coefficient of (C2), $k_\mathrm{C2, r}$, from \citet[][for low-pressure limit]{Page89_kinetics} and \citet[][for high-pressure limit]{Bezard02_carbon}, and the equilibrium constant, $K_\mathrm{C6, eq}$, of:
\renewcommand{\reactionprefix}{C}
\setcounter{reaction}{5}
\par\noindent
\begin{tabular*}{\columnwidth}{@{}r@{\;}l@{\extracolsep{\fill}}r@{}}
    \ce{CH3O} & \ce{<=> CO + 3/2 H2} & \reactiontag\label{path_Cooper2006_Keq} \\
\end{tabular*}
from \citet{Tsang86_chemical}.
Based on the above, \citet{Cooper06_dynamics} expressed the \ce{CO}-to-\ce{CH4} conversion timescale, $\tau_{\ce{CO}}$, as:
\begin{equation}
    \tau_{\ce{CO}}(p, T)=\frac{K_\mathrm{C6, eq}k_\mathrm{B}T}{k_\mathrm{C2, r}[\ce{H2}]^{3/2}p^{5/2}},
\end{equation}
where $p$ is pressure, $T$ is temperature, $k_\mathrm{B}$ is Boltzmann’s constant, and [\ce{H2}] is \ce{H2} mole fraction.
They then used this timescale to calculate the change in \ce{CO} mole fraction, [\ce{CO}], over time using the equation:
\begin{equation}
    \frac{D[\ce{CO}]}{Dt}=\frac{[\ce{CO}]_\mathrm{eq}-[\ce{CO}]}{\tau_{\ce{CO}}},
\end{equation}
where $[\ce{CO}]_\mathrm{eq}$ is \ce{CO} mole fraction at chemical equilibrium.
This equation states that if $[\ce{CO}]<[\ce{CO}]_\mathrm{eq}$, then \ce{CO} is produced, whereas if $[\ce{CO}]>[\ce{CO}]_\mathrm{eq}$, \ce{CO} is destroyed, and $\tau_{\ce{CO}}$ regulates the time it takes for such production or loss to occur.
In other words, \ce{CO} abundance is driven towards, or relaxed to, its equilibrium abundance over a prescribed chemical timescale.
This timescale changes as a function of pressure and temperature, and assumes there is a single \ce{CO}-to-\ce{CH4} conversion pathway controlling it.

The \citet{Cooper06_dynamics} chemical relaxation scheme is one of the simplest and computationally efficient methods for representing disequilibrium chemistry in hot-Jupiter GCMs; however, it neglects two important considerations.
First, it is more realistic to assume that the rate of change of a chemical species’ abundance depends on the net difference between its production and loss rates, rather than on the deviation of its current abundance from equilibrium.
Second, the production and loss rates of a species generally depend on multiple reactions in which it participates, rather than on a single rate-limiting reaction.
Both of these limitations are addressed if the temporal evolution of a species abundance is written as:
\begin{equation}
    \frac{Dn_i}{Dt} =
    \sum_j P_{ij}(\mathbf{n},p,T) - \sum_k L_{ik}(\mathbf{n},p,T),
\end{equation}
where $P_{ij}$ is the production rate of species $i$ from reaction $j$, $L_{ik}$ is the loss rate of species $i$ from reaction $k$, and $\mathbf{n}=(n_1, n_2,\ldots, n_N)$ is the vector of abundances of all species.
This equation relies on the rate laws of mass-action chemical kinetics and provides a more physically consistent description of disequilibrium chemistry.
It also makes chemical kinetics models a natural benchmark for validating simplified treatments of disequilibrium chemistry.

\citet{Tsai18_consistent} validated the \citet{Cooper06_dynamics} chemical relaxation scheme against 1D thermochemical kinetics simulations with the \texttt{VULCAN} \ce{C}-\ce{H}-\ce{O} chemical network \citep{Tsai17_vulcan}.
They found that the scheme is accurate to within about an order of magnitude, a factor of two, and 10\% for WASP-18b-, HD~189733b-, and GJ~1214b-like atmospheres, respectively.
The variation in the scheme's accuracy reflects the underlying complexity of carbon chemistry, with \citet{Tsai18_consistent} identifying three groups of \ce{CH4}-to-\ce{CO} conversion pathways relevant for hot Jupiters.
These pathways proceed
\begin{itemize}
    \item[(A)] through the progressive dehydrogenation of \ce{CH4} into \ce{C}, followed by oxidation into \ce{CO},
    \item[(B)] via \ce{H2CO} from directly oxidising \ce{CH3}, and
    \item[(C)] via intermediate species like \ce{CH2OH}, \ce{CH3OH}, or \ce{CH3O} from oxidising \ce{CH3}.
\end{itemize}
The rate-limiting steps for these pathways are:
\renewcommand{\reactionprefix}{T}
\setcounter{reaction}{0}
\par\noindent
\begin{tabular*}{\columnwidth}{@{}r@{\;}l@{\extracolsep{\fill}}r@{}}
    \ce{CH3 + OH + M} & \ce{-> CH3OH + M} & \reactiontag \\
    \ce{CH3 + OH} & \ce{-> CH2OH + H} & \reactiontag \\
    \ce{CH2OH + M} & \ce{-> H2CO + H + M} & \reactiontag \\
    \ce{CH3 + O(^{3}P)} & \ce{-> H2CO + H} & \reactiontag \\
    \ce{C + OH} & \ce{-> CO + H} & \reactiontag \\
    \ce{CH4 + H} & \ce{-> CH3 + H2} & \reactiontag \\
    \ce{CH3OH + H} & \ce{-> CH3O + H2} & \reactiontag \\
    \ce{C2H2 + O} & \ce{-> CH2 + CO} & \reactiontag \\
    \ce{CH3 + H2O} & \ce{-> CH3OH + H} & \reactiontag \\
    \ce{C2H2 + OH} & \ce{-> CH3 + CO} & \reactiontag \\
\end{tabular*}
Notably, none of these ten steps matches reaction~(C2), which \citet{Cooper06_dynamics} adopted as rate-limiting.
Using the later ab initio rate coefficients of \citet{Moses11_disequilibrium}, \citet{Tsai18_consistent} found that the subsequent reaction~(C3) is always slower than (C2), which makes (C3) the rate-limiting step.
The reverse of reaction~(C3) is reaction~(T7).

Chemical relaxation schemes have been used in several hot-Jupiter GCM studies.
The \citet{Cooper06_dynamics} scheme, besides being used in the \texttt{SPARC/MITgcm} GCM, was also implemented into the Met Office \texttt{Unified Model} (\texttt{UM}) GCM and applied to model the atmosphere of HD~209458b \citep{Drummond18_observable}.
\citet{Tsai18_consistent} proposed another, more generalised, chemical relaxation scheme, which was implemented into the \texttt{THOR} GCM and applied to model the atmosphere of WASP-43b \citep{Mendonca18_threedimensional}.
Other variants of chemical relaxation have also been used for ultra-hot Jupiters \citep[e.g.][]{Tan19_atmospheric}, though these studies apply relaxation to the thermal dissociation and recombination of \ce{H2} rather than to carbon chemistry.

With \textit{JWST} now delivering observations that demand more rigorous chemical modelling, the exoplanet community has begun transitioning to 3D chemical kinetics.
The \texttt{UM} GCM has two chemical kinetics schemes that work with either the \citet{Venot19_reduced} \ce{C}-\ce{H}-\ce{O}-\ce{N} reduced thermochemical network \citep{Drummond20_implications} or the \citet{Tsai22_minichemical} \ce{C}-\ce{H}-\ce{O}-\ce{N} \texttt{minichem} hybrid (combining elementary and net reactions) thermochemical network \citep{Christie26_benchmarking}.
\texttt{Exo-FMS} GCM employs the latter \citep{Lee23_minichemical}.
This indicates that solving relatively large chemical networks in hot-Jupiter GCMs has become computationally tractable.
However, relaxation schemes retain substantial practical advantages in 3D models \citep[e.g.][]{Steinrueck19_effect}, and 1D and pseudo-2D models remain the tool of choice for wide surveys of parameter space \citep[e.g.][]{Baeyens21_grid,Baeyens22_grid,Mukherjee24_sonora,Mukherjee25_effects}.

Chemical relaxation in GCMs and quench treatments in 1D models \citep[e.g.][]{Zahnle14_methane,Hu21_photochemistry,Mukherjee24_sonora,Mukherjee25_effects,Glein25_deciphering} share one assumption: that a single chemical timescale represents the conversion between \ce{CH4} and \ce{CO}.
Yet this assumption has not been tested against 3D chemical kinetics simulations of hot-Jupiter atmospheres.
This test is crucial for transport-induced quenching, because quenching arises from the competition between dynamical and chemical timescales.
\ce{CH4} provides a powerful tool for this test: quenching strongly affects \ce{CH4} abundance, and leaves clear observational signatures \citep{Zamyatina23_observability,Zamyatina24_quenchingdriven}. Moreover, accurate modelling of \ce{CH4} quenching is essential for inferring deep atmospheric temperatures \citep{Fortney20_beyond} and for correctly interpreting limb asymmetries \citep{Radica26_supersolar} and phase curves \citep{Bell24_nightside,Parmentier26_horizontal}.

This paper therefore asks three questions:
\begin{itemize}
    \item[(1)] Does the dominant \ce{CH4}-to-\ce{CO} conversion pathway vary at the \ce{CH4} quench level in a 3D atmosphere?
    \item[(2)] Does a single conversion timescale represent \ce{CH4}-to-\ce{CO} conversion at the \ce{CH4} quench level, as chemical relaxation assumes?
    \item[(3)] Does a 1D quench treatment hold in a 3D atmosphere?
\end{itemize}

We address these questions by analysing aerosol-free, solar-composition GCM simulations coupling hydrodynamics, radiative transfer and thermochemical kinetics (i.e., excluding photolysis) for the atmospheres of HAT-P-11b, HD~189733b, HD~209458b, and WASP-17b (Section~\ref{sec:methods}).
We find ten dominant \ce{CH4}-to-\ce{CO} conversion pathways in these atmospheres.
At the \ce{CH4} quench level, however, the dominant pathway varies only on WASP-17b, where two pathways compete, whereas a single pathway operates on HAT-P-11b, HD~189733b and HD~209458b (Sections~\ref{sec:results_equator} and~\ref{sec:results_3d}).
A single conversion timescale therefore represents \ce{CH4}-to-\ce{CO} conversion at the \ce{CH4} quench level on the three cooler planets, but not on WASP-17b.
This implies that chemical relaxation schemes are adequate for HAT-P-11b-, HD~189733b-, and HD~209458b-like atmospheres, but are inadequate for WASP-17b-like atmospheres (at least within the constraints of our GCM).
A 1D quench treatment built on the \citet{Zahnle14_methane} timescale holds only in part (Section~\ref{sec:results_zahnle}): it typically locates the \ce{CH4} quench level to within a factor of two in pressure on every planet, but holding abundances vertically uniform above the quench level misestimates the minor carbon carrier by up to two orders of magnitude.
Section~\ref{sec:results_zahnle_abundances} draws out the implications for both simplifications, Section~\ref{sec:discussion} discusses what quenching is, the uncertainties in modelling it in 3D, and its use as a diagnostic of atmospheric processes, and Section~\ref{sec:conclusions} summarises our conclusions.
Appendices~\ref{sec:appendix_quench_slices} and~\ref{sec:appendix_zahnle_slices} present the pressure--latitude structure behind the quench level analysis and the pressure--latitude comparison with \citet{Zahnle14_methane}.

\section{Methods}
\label{sec:methods}

To answer these questions, we first introduce our four target planets (Section~\ref{sec:targets}) and the GCM simulations we analyse (Section~\ref{sec:model}), and then describe how we diagnose quenching in these simulations (Section~\ref{sec:diagnosing_quenching}).

\subsection{The target planets}
\label{sec:targets}

Our four planets span nearly \SI{900}{\kelvin} in equilibrium temperature, from the warm super-Neptune HAT-P-11b \citep{Bakos10_hatp11b} --- the smallest and coldest of our planets, with a radius of $0.434\,R_\mathrm{J}$, a semimajor axis of $0.05325$~au, and an equilibrium temperature of \SI{869}{\kelvin} --- through the hot Jupiters HD~189733b \citep{Bouchy05_elodie} ($T_\mathrm{eq}=\SI{1230}{\kelvin}$, $R_\mathrm{p}=1.154\,R_\mathrm{J}$, $a=0.02984$~au) and HD~209458b \citep{Henry99_transiting} ($T_\mathrm{eq}=\SI{1459}{\kelvin}$, $R_\mathrm{p}=1.380\,R_\mathrm{J}$, $a=0.04747$~au), to the highly inflated hot Jupiter WASP-17b \citep{Anderson10_wasp17b} --- the largest and hottest of our planets, with a radius of $1.932\,R_\mathrm{J}$, a semimajor axis of $0.05135$~au, and an equilibrium temperature of \SI{1755}{\kelvin}.
Full stellar and planetary parameters, primarily sourced from the TEPCat data base \citep{Southworth11_homogeneous}, are provided in \citet{Zamyatina23_observability}.

\subsection[The Unified Model]{The \texttt{Unified Model}}
\label{sec:model}

We analyse the Met Office \texttt{Unified Model} (\texttt{UM}) GCM simulations with thermochemical kinetics presented by \citet{Drummond20_implications} and \citet{Zamyatina23_observability}.
In brief, the model's dynamical core \citep[\texttt{ENDGame},][]{Wood14_inherently,Mayne14_using,Mayne14_unified} solves the full, deep-atmosphere, non-hydrostatic Euler equations on a constant latitude--longitude grid using a semi-implicit, semi-Lagrangian advection scheme.
The radiative transfer component \citep[\texttt{Socrates},][]{Edwards96_efficient,Edwards96_studies,Amundsen14_accuracy,Amundsen16_uk,Amundsen17_treatment} solves the two-stream equations and treats opacities using the correlated-\textit{k} and equivalent extinction methods.
The incoming stellar radiation is attenuated in the plane-parallel approximation, along the vertical column scaled by the stellar zenith angle, so shortwave heating falls to zero at the terminators; the \texttt{UM}'s pseudo-spherical irradiation scheme \citep{Jackson20_space,Christie22_impact}, which lights the twilight region beyond the terminator, is not used.
Opacity sources include absorption by \ce{H2O}, \ce{CO}, \ce{CO2}, \ce{CH4}, \ce{NH3}, \ce{HCN}, \ce{Li}, \ce{Na}, \ce{K}, \ce{Rb}, and \ce{Cs}, collision-induced absorption from \ce{H2}-\ce{H2} and \ce{H2}-\ce{He}, and Rayleigh scattering by \ce{H2} and \ce{He} \citep[][Appendix A]{Zamyatina23_observability}.
The chemistry component solves stiff ordinary differential equations describing production and loss of \ce{^{1}CH2}, \ce{^{3}CH2}, \ce{CH2OH}, \ce{CH3}, \ce{CH3O}, \ce{CH3OH}, \ce{CH4}, \ce{CN}, \ce{CO}, \ce{CO2}, \ce{H}, \ce{H2}, \ce{H2CN}, \ce{H2CO}, \ce{H2O}, \ce{HCO}, \ce{HCN}, \ce{HNCO}, \ce{HOCN}, \ce{N2}, \ce{N2H2}, \ce{N2H3}, \ce{NCO}, \ce{NH}, \ce{NH2}, \ce{NH3}, \ce{NNH}, \ce{O(^{3}P)}, and \ce{OH} included in the \citet{Venot19_reduced} \ce{C}-\ce{H}-\ce{O}-\ce{N} reduced thermochemical network, with \ce{He} being involved only as a third body.
\ce{C2H2} is not part of this network, so the \citet{Tsai18_consistent} steps (T8) and (T10), which involve \ce{C2H2}, cannot occur.
Alkali species, which are also not part of the network, are parametrised using the threshold method of \citet{Amundsen16_uk}.
The network contains no photolysis reactions, so our simulations include no photochemistry.
The chemistry solver stops integrating for the rest of a chemical time-step once a numerical escape criterion is met.
\citet{Christie26_benchmarking} show that this criterion causes artificial quenching: on WASP-96b, a planet not analysed here, it raises the quench pressure and overestimates \ce{CH4} by factors of 1.5 to 3.
Our simulations use this criterion, so our planets may be affected to an unknown degree.
All simulations assume aerosol-free atmospheres with solar metallicity and a \ce{C}/\ce{O} ratio of 0.55 \citep{Asplund09_chemical}.
Our grid is too coarse to resolve convective plumes, and we apply no dry convective adjustment scheme, so vertical mixing is handled only by the resolved dynamics of the non-hydrostatic core.
Chemical species are carried as tracers and transported by the same advection scheme as the other prognostic fields, so their vertical transport, and hence advection timescales, derive entirely from the resolved large-scale flow.

\subsection[Diagnosing quenching]{Diagnosing quenching}
\label{sec:diagnosing_quenching}

To diagnose quenching in our simulations, we first define quenching (Section~\ref{sec:def_quenching}), then calculate the advection (Section~\ref{sec:calc_tau_adv}) and \ce{CH4}-to-\ce{CO} conversion timescales (Section~\ref{sec:calc_tau_chem}), and finally identify the \ce{CH4} quench level and quench region from the ratio of the two timescales (Section~\ref{sec:calc_ch4_qn_level}).\footnote{We calculate both timescales and the quench level with the \texttt{quenchpath} Python package \citep{quenchpath}.}

\subsubsection{Definition of quenching}
\label{sec:def_quenching}

Quenching is the condition at which control of a species' abundance passes from thermochemistry to transport.
Below the quench level the reactions that interconvert the species are faster than transport, and the species' abundance is set by local temperature and pressure.
At the quench level the interconverting reactions and transport are equally fast.
Above the quench level the interconverting reactions are slower than transport, and the species' abundance is governed by transport together with the slower but still active conversion chemistry.
Quenching is therefore localised to the transition itself, occurring at the quench level rather than persisting as a state of the species thereafter.
We use the term \textit{quenching} only for that transition, and refer to abundances above the quench level by position rather than as quenched.
This usage deliberately departs from the common association of quenching with a well-mixed profile (Section~\ref{sec:disc_quenching}).

Susceptibility to quenching is a property of a species together with the interconversion considered, not of the species alone: the condition is met only where the reactions driving that interconversion can become as slow as transport.
A species may quench with respect to one interconversion, yet the abundance of that species may still be modified above the quench level by photochemistry, aerosol formation, or advection from a region of differing abundance.

\subsubsection{Calculation of advection timescales}
\label{sec:calc_tau_adv}

To understand where transport becomes relevant to the chemical evolution, we compute the zonal ($\tau^{\mathrm{u}}_{\mathrm{adv}}$), meridional ($\tau^{\mathrm{v}}_{\mathrm{adv}}$), and vertical ($\tau^{\mathrm{w}}_{\mathrm{adv}}$) advection timescales using Equations 7--9 from \citet{Drummond18_observable}:
\begin{equation}
    \tau^{\mathrm{u}}_{\mathrm{adv}}=\frac{2 \pi R_\mathrm{p}}{|\bar{u}|}, \quad
    \tau^{\mathrm{v}}_{\mathrm{adv}}=\frac{\pi R_\mathrm{p}}{2 |\bar{v}|}, \quad
    \tau^{\mathrm{w}}_{\mathrm{adv}}=\frac{H}{|\bar{w}|},
\end{equation}
where $R_\mathrm{p}$ is the planetary radius, $\bar{u}$ is the zonal wind speed averaged over the dayside (\ang{90}E $< \lambda <$ \ang{270}E) and over all latitudes, $\bar{v}$ is the meridional wind speed averaged over the dayside and over the northern hemisphere (\ang{0} $< \phi <$ \ang{90}N), and $\bar{w}$ and $H$ are the vertical wind speed and the scale height, both averaged over the dayside and over all latitudes.
We also compute the dominant, i.e., the shortest, advection timescale, $\tau^{\mathrm{dom}}_{\mathrm{adv}}$, as:
\begin{equation}
    \tau^{\mathrm{dom}}_{\mathrm{adv}}=\min(\tau^{\mathrm{u}}_{\mathrm{adv}}, \tau^{\mathrm{v}}_{\mathrm{adv}}, \tau^{\mathrm{w}}_{\mathrm{adv}}).
\end{equation}
$\tau^{\mathrm{dom}}_{\mathrm{adv}}$ represents the most efficient direction of atmospheric transport at each height, and is used to define the \ce{CH4} quench level by comparison with the dominant \ce{CH4}-to-\ce{CO} conversion timescale $\tau^{\mathrm{dom}}_{\ce{CH4-to-CO}}$ (Section~\ref{sec:calc_tau_chem}).

\subsubsection[Calculation of CH4-to-CO conversion timescales]{Calculation of CH\textsubscript{4}-to-CO conversion timescales}
\label{sec:calc_tau_chem}

In each model grid cell, \ce{CH4} can be converted to \ce{CO} via multiple reaction pathways. In this work, however, we approximate the total conversion rate by the rate along the dominant, i.e., the fastest, pathway.
To identify this pathway, we first evaluate reaction rates in each model grid cell at the final simulation timestep.
Then, using the \texttt{networkx} Python package \citep{networkx}, we construct a directed graph of the full network, with chemical species as nodes, reactions as edges, and inverse reaction rates as edge weights\footnote{
Graph algorithms require edge weights representing a cost. For a chemical reaction, that cost is the time required per molecule\,\si{\per\centi\metre\cubed} converted. Faster reactions therefore carry smaller edge weights.}.
Each reversible reaction is oriented in the direction favoured thermodynamically by its equilibrium constant.
The dominant pathway, i.e., the one with the smallest sum of edge weights, is identified using Dijkstra's algorithm \citep{Dijkstra59_note}, applied to the subgraph containing only C-bearing species.
We perform this analysis in every model grid cell, at the model's native grid resolution (144 longitudes, 90 latitudes, 66 height levels).
The dominant \ce{CH4}-to-\ce{CO} conversion timescale, $\tau^{\mathrm{dom}}_{\ce{CH4-to-CO}}$, is computed as:
\begin{equation}
    \tau^{\mathrm{dom}}_{\ce{CH4-to-CO}}=\frac{n_{\ce{CH4}}}{R_{\ce{CH4-to-CO}}},
\end{equation}
where $n_{\ce{CH4}}$ is \ce{CH4} number density, and $R_{\ce{CH4-to-CO}}$ is the rate of the dominant \ce{CH4}-to-\ce{CO} conversion pathway, i.e., the inverse of the sum of its edge weights.
Notably, $R_{\ce{CH4-to-CO}}$ is the rate of the full pathway rather than of its rate-limiting step.
Each step adds its own timescale to that of the pathway, so using the rate-limiting step alone would produce a shorter conversion timescale and a lower \ce{CH4} quench pressure.

\subsubsection[Identification of the CH4 quench level and quench region]{Identification of the CH\textsubscript{4} quench level and quench region}
\label{sec:calc_ch4_qn_level}

The quench level lies where conversion and transport are equally fast.
Thermodynamics sets the direction of conversion, which at the quench level on all of our four planets is towards \ce{CO}.
We therefore define the \ce{CH4} quench level as the level at which the dominant \ce{CH4}-to-\ce{CO} conversion timescale equals the dominant advection timescale:
\begin{equation}
\label{eq:quench_level}
    \frac{\tau^{\mathrm{dom}}_{\ce{CH4-to-CO}}}{\tau^{\mathrm{dom}}_{\mathrm{adv}}} = 1.
\end{equation}
Because $\tau^{\mathrm{dom}}_{\ce{CH4-to-CO}}$ is evaluated locally, the quench level can shift with local temperature and pressure, and wherever the dominant pathway changes.
This timescale equality corresponds to a Damk\"{o}hler number of unity \citep{Damkohler36_einflusse}.
In combustion science and chemical engineering, the Damk\"{o}hler number, a ratio of transport to chemical timescales, separates chemistry- and transport-dominated regimes.
In reality the transition between these regimes is gradual, so the \ce{CH4} quench level defined by Equation~(\ref{eq:quench_level}) only approximates where \ce{CH4} departs from thermochemical equilibrium.

We therefore additionally define the \ce{CH4} quench region as the band around the \ce{CH4} quench level within which
\begin{equation}
\label{eq:quench_region}
    \left| \ln\left(\frac{\tau^{\mathrm{dom}}_{\ce{CH4-to-CO}}}{\tau^{\mathrm{dom}}_{\mathrm{adv}}}\right) \right| \leq b,
\end{equation}
\noindent where $b$ is the half-width of the quench region in e-foldings, so that the quench region spans timescale ratios between $e^{-b}$ and $e^{b}$.
We take $b=1$, that is, the two timescales are within a factor of $e$ of one another, \numrange{0.37}{2.72}.
The logarithm itself is zero at the quench level, negative below the quench level, where conversion outpaces transport, and positive above the quench level, where transport outpaces conversion.

\begin{figure*}
    \centering
    \includegraphics[width=\textwidth]{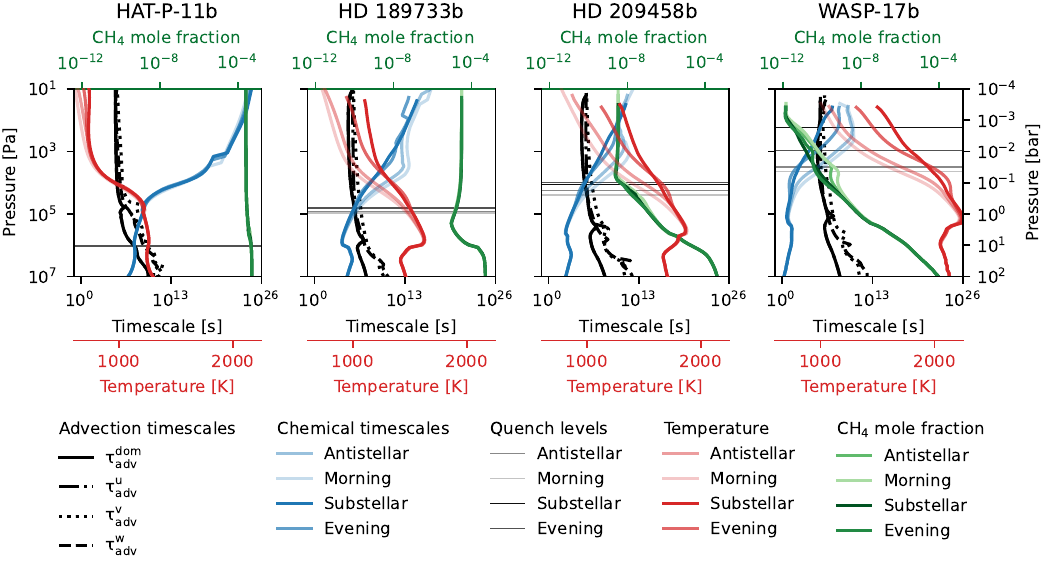}
    \caption{
    Equatorial vertical profiles of the dominant \ce{CH4}-to-\ce{CO} conversion timescale ($\tau^{\mathrm{dom}}_{\ce{CH4-to-CO}}$), the gas temperature and the \ce{CH4} mole fraction at the antistellar, morning, substellar, and evening longitudes of HAT-P-11b, HD~189733b, HD~209458b, and WASP-17b.
    Black lines show the zonal ($\tau^{\mathrm{u}}_{\mathrm{adv}}$), meridional ($\tau^{\mathrm{v}}_{\mathrm{adv}}$), vertical ($\tau^{\mathrm{w}}_{\mathrm{adv}}$) and dominant ($\tau^{\mathrm{dom}}_{\mathrm{adv}}$) advection timescales, which are dayside averages and so do not vary with longitude (Section~\ref{sec:calc_tau_adv}).
    Horizontal lines mark the \ce{CH4} quench level, shaded by longitude.
    }
    \label{fig:pres_vp_taus}
\end{figure*}

\section{Results}
\label{sec:results}

To build intuition, we first examine \ce{CH4} quenching and abundance at the equator, using vertical profiles at the antistellar, morning, substellar, and evening longitudes (Section~\ref{sec:results_equator}), and then in 3D, using pressure--latitude slices at the same longitudes (Section~\ref{sec:results_3d}).
We then test a 1D quench treatment built on the analytic timescale of \citet{Zahnle14_methane} against our 3D simulations (Section~\ref{sec:results_zahnle}), and finally draw out what the results imply for chemical relaxation in GCMs and 1D quench schemes (Section~\ref{sec:results_zahnle_abundances}).

\subsection[CH4 quenching and abundance at the equator]{CH\textsubscript{4} quenching and abundance at the equator}
\label{sec:results_equator}

The advection directions that compete with the conversion chemistry at the \ce{CH4} quench level vary across our four planets (Figure~\ref{fig:pres_vp_taus}).
On HAT-P-11b, $\tau^{\mathrm{u}}_{\mathrm{adv}}$ is shorter than $\tau^{\mathrm{v}}_{\mathrm{adv}}$ and $\tau^{\mathrm{w}}_{\mathrm{adv}}$ by more than an order of magnitude, so zonal advection sets the quench level.
On the other three planets, $\tau^{\mathrm{u}}_{\mathrm{adv}}$ and $\tau^{\mathrm{w}}_{\mathrm{adv}}$ lie within a factor of three of one another, while $\tau^{\mathrm{v}}_{\mathrm{adv}}$ is longer than both, by up to an order of magnitude.
Zonal and vertical advection therefore both compete with the conversion chemistry at the quench level of these three planets, and meridional advection does not.

\ce{CH4} quenches deeper in colder atmospheres.
On HAT-P-11b, the \ce{CH4} quench pressure is $\approx$\SI{e6}{\pascal} across the entire equator.
On HD~189733b, it is $\approx$\SI{9e4}{\pascal} at the colder antistellar and morning longitudes, and $\approx$\SI{7e4}{\pascal} at the hotter substellar and evening longitudes.
\ce{CH4} behaves similarly on HD~209458b, quenching at $\approx$\SI{2e4}{\pascal} at the antistellar and morning longitudes and at $\approx$\SI{1e4}{\pascal} at the substellar and evening longitudes.
On WASP-17b, however, \ce{CH4} quenching is more spatially variable, occurring at \SIrange{3e3}{4e3}{\pascal} at the antistellar and morning longitudes, shifting to $\approx$\SI{2e2}{\pascal} at the substellar longitude, and deepening to $\approx$\SI{9e2}{\pascal} at the evening longitude.

Furthermore, \ce{CH4} quenches hotter as well as shallower in hotter atmospheres, so the quench temperature differs between planets: at $\approx$\SI{1250}{\kelvin} on HAT-P-11b, $\approx$\SI{1500}{\kelvin} on HD~189733b, $\approx$\SI{1600}{\kelvin} on HD~209458b and $\approx$\SI{1700}{\kelvin} on WASP-17b.

Finally, because \ce{CH4} quenches shallower and hotter in hotter atmospheres, where \ce{CH4} is less abundant in thermochemical equilibrium, the \ce{CH4} mole fraction at the quench level falls across our four planets.
That mole fraction is $\approx$\num{5e-4} on HAT-P-11b, $\approx$\num{1.5e-5} on HD~189733b, $\approx$\num{1e-8} on HD~209458b and \numrange{3e-12}{7e-10} on WASP-17b.
Above the quench level, however, the \ce{CH4} mole fraction keeps changing with height.
By the top of the model domain, the \ce{CH4} mole fraction has fallen to $\approx$\num{3e-4} on HAT-P-11b, $\approx$\num{3e-9} on HD~209458b and $\approx$\num{1e-12} on WASP-17b, and risen to $\approx$\num{3e-5} on HD~189733b, at all four longitudes.
Holding \ce{CH4} vertically uniform above the quench level would therefore misestimate \ce{CH4} on all four planets.

\subsection[CH4 quenching and abundance in 3D]{CH\textsubscript{4} quenching and abundance in 3D}
\label{sec:results_3d}

To explain why the equatorial \ce{CH4} quench level shifts with longitude and why \ce{CH4} abundance keeps changing above the quench level, we extend the analysis to 3D.
The quench level depends on the local temperature and pressure and on which \ce{CH4}-to-\ce{CO} pathway dominates in each grid cell (Section~\ref{sec:calc_ch4_qn_level}).
We therefore first identify where each pathway dominates (Section~\ref{sec:results_3d_paths}), then examine what shapes the quench level in 3D (Section~\ref{sec:results_3d_mechanism}), and finally what shapes \ce{CH4} abundance above the quench level (Section~\ref{sec:results_3d_above}).

\subsubsection[Spatial variability of CH4-to-CO conversion pathways]{Spatial variability of CH\textsubscript{4}-to-CO conversion pathways}
\label{sec:results_3d_paths}

We find ten dominant \ce{CH4}-to-\ce{CO} conversion pathways across our four planets (Table~\ref{tab:pathways}).

\begin{table}
\centering
\caption{The dominant \ce{CH4}-to-\ce{CO} conversion pathways found in this work.}
\label{tab:pathways}
\footnotesize\setlength{\tabcolsep}{4pt}
\begin{tabular}{r@{\;}lc}
\hline
\multicolumn{2}{c}{Pathway} & Index \\
\hline
\ce{CH4 + H} & \ce{-> CH3 + H2} & \multirow{6}{*}{\refstepcounter{pathway}\label{pathway1}(\thepathway)} \\
\ce{CH3 + OH ( + M)} & \ce{-> CH3OH ( + M)} & \\
\ce{CH3OH + H} & \ce{-> CH2OH + H2} & \\
\ce{CH2OH + M} & \ce{-> H2CO + H + M} & \\
\ce{H2CO + H} & \ce{-> HCO + H2} & \\
\ce{HCO + M} & \ce{-> CO + H + M} & \\
\hline
\ce{CH4 + H} & \ce{-> CH3 + H2} & \multirow{6}{*}{\refstepcounter{pathway}\label{pathway2}(\thepathway)} \\
\ce{CH3 + OH ( + M)} & \ce{-> CH3OH ( + M)} & \\
\ce{CH3OH + H} & \ce{-> CH3O + H2} & \\
\ce{CH3O + M} & \ce{-> H2CO + H + M} & \\
\ce{H2CO + H} & \ce{-> HCO + H2} & \\
\ce{HCO + M} & \ce{-> CO + H + M} & \\
\hline
\ce{CH4 + H} & \ce{-> CH3 + H2} & \multirow{7}{*}{\refstepcounter{pathway}\label{pathway3}(\thepathway)} \\
\ce{CH3 + OH} & \ce{-> ^{1}CH2 + H2O} & \\
\ce{^{1}CH2 + H2O} & \ce{-> CH3OH} & \\
\ce{CH3OH + H} & \ce{-> CH3O + H2} & \\
\ce{CH3O + M} & \ce{-> H2CO + H + M} & \\
\ce{H2CO + H} & \ce{-> HCO + H2} & \\
\ce{HCO + M} & \ce{-> CO + H + M} & \\
\hline
\ce{CH4 + H} & \ce{-> CH3 + H2} & \multirow{4}{*}{\refstepcounter{pathway}\label{pathway4}(\thepathway)} \\
\ce{CH3 + O(^{3}P)} & \ce{-> H2CO + H} & \\
\ce{H2CO + H} & \ce{-> HCO + H2} & \\
\ce{HCO + M} & \ce{-> CO + H + M} & \\
\hline
\ce{CH4 + H} & \ce{-> CH3 + H2} & \multirow{6}{*}{\refstepcounter{pathway}\label{pathway5}(\thepathway)} \\
\ce{CH3 + OH ( + M)} & \ce{-> CH3OH ( + M)} & \\
\ce{CH3OH + H} & \ce{-> CH2OH + H2} & \\
\ce{CH2OH + H} & \ce{-> H2CO + H2} & \\
\ce{H2CO + H} & \ce{-> HCO + H2} & \\
\ce{HCO + H} & \ce{-> CO + H2} & \\
\hline
\ce{CH4 + H} & \ce{-> CH3 + H2} & \multirow{7}{*}{\refstepcounter{pathway}\label{pathway6}(\thepathway)} \\
\ce{CH3 + OH} & \ce{-> ^{1}CH2 + H2O} & \\
\ce{^{1}CH2 + H2O} & \ce{-> CH3OH} & \\
\ce{CH3OH + H} & \ce{-> CH2OH + H2} & \\
\ce{CH2OH + H} & \ce{-> H2CO + H2} & \\
\ce{H2CO + H} & \ce{-> HCO + H2} & \\
\ce{HCO + H} & \ce{-> CO + H2} & \\
\hline
\ce{CH4 + H} & \ce{-> CH3 + H2} & \multirow{6}{*}{\refstepcounter{pathway}\label{pathway7}(\thepathway)} \\
\ce{CH3 + OH} & \ce{-> ^{1}CH2 + H2O} & \\
\ce{^{1}CH2 + M} & \ce{-> ^{3}CH2 + M} & \\
\ce{^{3}CH2 + CO2} & \ce{-> H2CO + CO} & \\
\ce{H2CO + H} & \ce{-> HCO + H2} & \\
\ce{HCO + H} & \ce{-> CO + H2} & \\
\hline
\ce{CH4 + H} & \ce{-> CH3 + H2} & \multirow{7}{*}{\refstepcounter{pathway}\label{pathway8}(\thepathway)} \\
\ce{CH3 + OH} & \ce{-> ^{1}CH2 + H2O} & \\
\ce{^{1}CH2 + H2O} & \ce{-> CH3OH} & \\
\ce{CH3OH + H} & \ce{-> CH3O + H2} & \\
\ce{CH3O + M} & \ce{-> H2CO + H + M} & \\
\ce{H2CO + H} & \ce{-> HCO + H2} & \\
\ce{HCO + H} & \ce{-> CO + H2} & \\
\hline
\ce{CH4 + H} & \ce{-> CH3 + H2} & \multirow{7}{*}{\refstepcounter{pathway}\label{pathway9}(\thepathway)} \\
\ce{CH3 + OH} & \ce{-> ^{1}CH2 + H2O} & \\
\ce{^{1}CH2 + H2O} & \ce{-> CH3OH} & \\
\ce{CH3OH + H} & \ce{-> CH2OH + H2} & \\
\ce{CH2OH + H} & \ce{-> H2CO + H2} & \\
\ce{H2CO + H} & \ce{-> HCO + H2} & \\
\ce{HCO + M} & \ce{-> CO + H + M} & \\
\hline
\ce{CH4 + H} & \ce{-> CH3 + H2} & \multirow{6}{*}{\refstepcounter{pathway}\label{pathway10}(\thepathway)} \\
\ce{CH3 + OH ( + M)} & \ce{-> CH3OH ( + M)} & \\
\ce{CH3OH + H} & \ce{-> CH2OH + H2} & \\
\ce{CH2OH + H} & \ce{-> H2CO + H2} & \\
\ce{H2CO + H} & \ce{-> HCO + H2} & \\
\ce{HCO + M} & \ce{-> CO + H + M} & \\
\hline
\end{tabular}
\end{table}

All our pathways start with the same reaction, the abstraction of \ce{H} from \ce{CH4} to form \ce{CH3}.
After that the pathways diverge, though some share steps.
\begin{itemize}
\item In pathway~(\ref{pathway4}), \ce{CH3} reacts with \ce{O(^{3}P)} to produce \ce{H2CO}.
\ce{H2CO} then reacts with \ce{H} to form \ce{HCO}, which then dissociates into \ce{CO} and \ce{H}.
Pathway~(\ref{pathway4}) matches the example pathway for group (B) in \citet{Tsai18_consistent}, except for the final reaction.
\item In all our other pathways, \ce{CH3} reacts with \ce{OH}.
Subsequent conversion to \ce{CO} involves \ce{CH3OH} in every case except pathway~(\ref{pathway7}).
\item Pathways~(\ref{pathway2}), (\ref{pathway3}) and (\ref{pathway8}) convert \ce{CH3OH} to \ce{CO} via \ce{CH3O}.
These three pathways differ in how \ce{CH3OH} is produced: in a three-body reaction or two bimolecular reactions involving \ce{^{1}CH2}, and in how \ce{HCO} is destroyed: by dissociation or \ce{H} abstraction.
Pathway~(\ref{pathway2}) matches the example pathway for group (C) in \citet{Tsai18_consistent}.
\item Pathways~(\ref{pathway1}), (\ref{pathway5}), (\ref{pathway6}), (\ref{pathway9}) and (\ref{pathway10}) convert \ce{CH3OH} to \ce{CO} via \ce{CH2OH}.
Pathway~(\ref{pathway1}), the most common of all our pathways, is the only one in which \ce{CH2OH} dissociates in a three-body reaction rather than by \ce{H} abstraction.
These five pathways also differ in how \ce{CH3OH} is produced: in a three-body reaction or two bimolecular reactions involving \ce{^{1}CH2}, and in how \ce{HCO} is destroyed: by dissociation or \ce{H} abstraction.
\item Pathway~(\ref{pathway7}), like pathway~(\ref{pathway4}), bypasses creation of single carbon-oxygen bonds.
\ce{CH3} transforms into \ce{^{1}CH2}, and then into \ce{^{3}CH2}.
Next, \ce{^{3}CH2} abstracts a double bonded oxygen from \ce{CO2} to form \ce{H2CO} \citep{Kovacs01_ch2}\footnote{The step \ce{^{3}CH2 + CO2 -> H2CO + CO} is spin-forbidden, since its triplet reactants form singlet products.
Its rate coefficient in our network derives from \citet{Tsang86_chemical}, who adopt the room-temperature flash-photolysis measurement of \citet{Laufer77_reaction}.
\citet{Kovacs01_ch2} conclude that the reaction is unlikely to proceed on the triplet surface, and that excitation to the singlet state is thermochemically favoured.
Pathway~(\ref{pathway7}) may therefore not be physically meaningful, but it does not operate at the \ce{CH4} quench level on any of our planets, so our conclusions do not depend on it.}.
Lastly, \ce{H2CO} loses an \ce{H} to form \ce{HCO}, which is then destroyed by \ce{H} abstraction.
\end{itemize}
\begin{figure*}
    \centering
    \includegraphics[width=\textwidth]{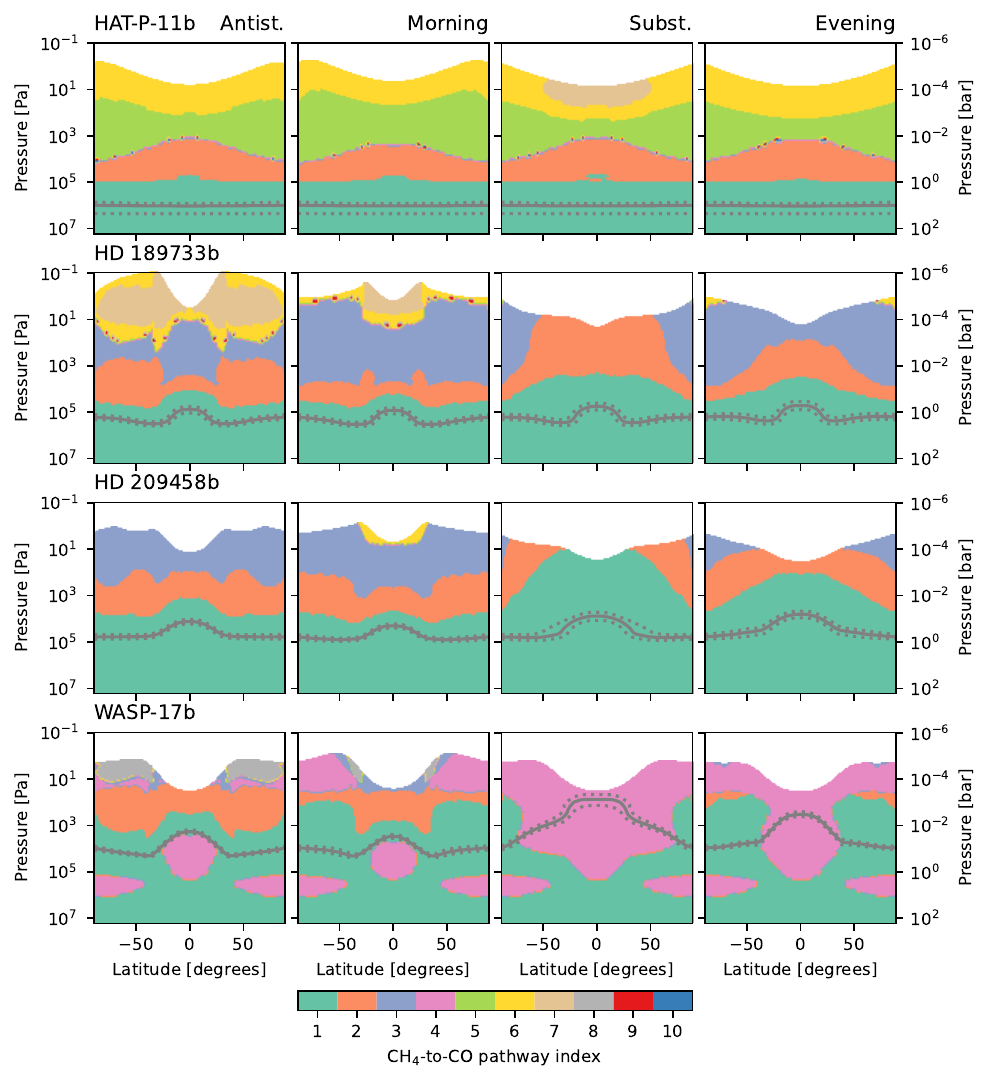}
    \caption{
    Pressure--latitude slices of the dominant \ce{CH4}-to-\ce{CO} conversion pathways
    at the antistellar, morning, substellar, and evening longitudes of the atmospheres of HAT-P-11b, HD~189733b, HD~209458b, and WASP-17b.
    Pathways are numbered in order of the fraction of grid points at which each is dominant, pooled over all four planets.
    The \ce{CH4} quench level is shown by a solid grey line, and the \ce{CH4} quench region by dotted grey lines.
    }
    \label{fig:slices_pres_local_pathways}
\end{figure*}

Grouping the chemically similar pathways, those via \ce{CH3O} ((\ref{pathway2}), (\ref{pathway3}), (\ref{pathway8})) and those via \ce{CH2OH} ((\ref{pathway1}), (\ref{pathway5}), (\ref{pathway6}), (\ref{pathway9}), (\ref{pathway10})), leaves four distinct groups: these two, (\ref{pathway4}) and (\ref{pathway7}).
If we group our pathways into the same (A), (B) and (C) groups as \citet{Tsai18_consistent}, our pathway~(\ref{pathway4}) falls into group (B), while all our other pathways fall into group (C), with the exception of (\ref{pathway7}), which fits none of the three: its carbon is dehydrogenated only as far as \ce{^{3}CH2}, and is then oxidised by \ce{CO2} rather than by \ce{OH} or \ce{O(^{3}P)}.

Pathways~(\ref{pathway1}) and (\ref{pathway2}) are the only two found on all our planets, and differ mainly in depth (Figure~\ref{fig:slices_pres_local_pathways}).
Pathway~(\ref{pathway1}), the most common of all our pathways, occupies the deep atmosphere, and (\ref{pathway2}) the layers directly above it.
How far (\ref{pathway1}) reaches upwards varies between planets, from \SI{5e4}{\pascal} on HAT-P-11b to \SI{10}{\pascal} on HD~209458b.
The remaining pathways are localised and do not occur on all planets.
Pathway~(\ref{pathway4}) is confined to WASP-17b, where it is the second most common pathway after (\ref{pathway1}), forming a deep-atmosphere band and dominating the dayside higher up.
Pathways~(\ref{pathway5}) and (\ref{pathway10}) are confined to HAT-P-11b: (\ref{pathway5}) spans the mid-atmosphere, and (\ref{pathway10}) forms small pockets along the lower boundary of the (\ref{pathway5}) layer.
Pathways~(\ref{pathway3}) and (\ref{pathway6}) each occur on three planets: (\ref{pathway3}) on HD~189733b, HD~209458b and WASP-17b, and (\ref{pathway6}) on HAT-P-11b, HD~189733b and HD~209458b.
Of the four rarest pathways overall, (\ref{pathway7}) on HD~189733b, (\ref{pathway8}) in the upper atmosphere of WASP-17b, and (\ref{pathway9}) on HD~189733b and HD~209458b cluster on the nightside and the terminators and avoid substellar longitudes almost entirely.
On HAT-P-11b, by contrast, (\ref{pathway7}) covers the substellar and evening quadrants at low latitudes, and (\ref{pathway10}) shows no preference in longitude.
The \ce{CH4}-to-\ce{CO} conversion timescales (Figure~\ref{fig:slices_pres_local_tau_ch4_net}) span \SIrange{0.125}{2.52e25}{\second} across our four planets (from a fraction of a second to about \num{8e17} years), i.e., about 26 orders of magnitude.
For \ce{CH4} quenching, however, only the timescales within the quench region matter, and these span just \SIrange{4.07e4}{6.28e7}{\second} (from about \SI{11}{\hour} to two years), i.e., about 3 orders of magnitude.

\subsubsection[What shapes the CH4 quench level in 3D]{What shapes the CH\textsubscript{4} quench level in 3D}
\label{sec:results_3d_mechanism}

To identify what shapes the \ce{CH4} quench level in 3D, we examine pressure--latitude slices of the dominant pathways and \ce{CH4} mole fraction (Figures~\ref{fig:slices_pres_local_pathways} and~\ref{fig:slices_pres_local_ch4_molefrac}), of the underlying timescale ratio (Figure~\ref{fig:slices_pres_local_tau_ch4_net_over_tau_ls_dom}), and of the total wind speed and gas temperature (Figures~\ref{fig:slices_pres_local_wind} and~\ref{fig:slices_pres_local_temp}), each marked with the \ce{CH4} quench level (Equation~(\ref{eq:quench_level})) and quench region (Equation~(\ref{eq:quench_region})).
Our advection timescale is a dayside average over all latitudes and varies only with height (Section~\ref{sec:calc_tau_adv}).
Every spatial variation in the quench level is therefore a variation in $\tau^{\mathrm{dom}}_{\ce{CH4-to-CO}}$, which depends on the local temperature and pressure and on which pathway dominates.

The \ce{CH4} quench level varies in 3D because two effects shorten its conversion timescale, and so raise the quench level: (a) a hotter atmosphere and (b) a faster pathway.
Between them they account for the equatorial \ce{CH4} depletion on HD~189733b, HD~209458b and WASP-17b.
Where the atmosphere is hotter the conversion is faster, the conversion timescale is shorter, and the quench level sits higher; and because \ce{CH4} is less abundant at equilibrium higher up, a higher quench level means a lower quench-level abundance.
On HD~189733b and HD~209458b temperature acts alone, since a single pathway, (\ref{pathway1}), dominates at the quench level at every latitude.
On WASP-17b, however, the dominant pathway shortens the conversion timescale further.
Pathway~(\ref{pathway1}) contains two termolecular steps beyond the final dissociation it shares with (\ref{pathway4}), so its rate rises with pressure as well as with temperature.
Pathway~(\ref{pathway4}) contains no termolecular step beyond that one, but its second reaction consumes \ce{O(^{3}P)}, which is held in equilibrium with \ce{H2O} by fast H-atom exchange and whose abundance rises steeply with temperature.
Pathway~(\ref{pathway4}) therefore has the larger effective activation energy and the weaker pressure dependence, and takes over where the atmosphere is hot and the pressure low.

How far the dominant pathway switch moves WASP-17b's quench level depends on longitude.
The switch acts mainly at the substellar longitude, where the quench level forms a broad dome over the tropics and (\ref{pathway4}) dominates across most of it.
Relocating the quench level with (\ref{pathway1}) alone, holding the simulated atmosphere fixed (not shown), lowers the equatorial quench level and raises the equatorial \ce{CH4} abundance at the quench level by a factor of about four; at the other longitudes the quench level barely moves.
Pathway~(\ref{pathway4}) does not destroy \ce{CH4} faster than (\ref{pathway1}) would: below the quench level the abundance sits at its equilibrium value whichever pathway carries the conversion.
What the switch does is move the quench level.
Therefore, the equatorial \ce{CH4} depletion is amplified by the switch from (\ref{pathway1}) to (\ref{pathway4}) on WASP-17b.

\subsubsection[What shapes CH4 abundance above the quench level]{What shapes CH\textsubscript{4} abundance above the quench level}
\label{sec:results_3d_above}

\begin{figure*}
    \centering
    \includegraphics[width=\textwidth]{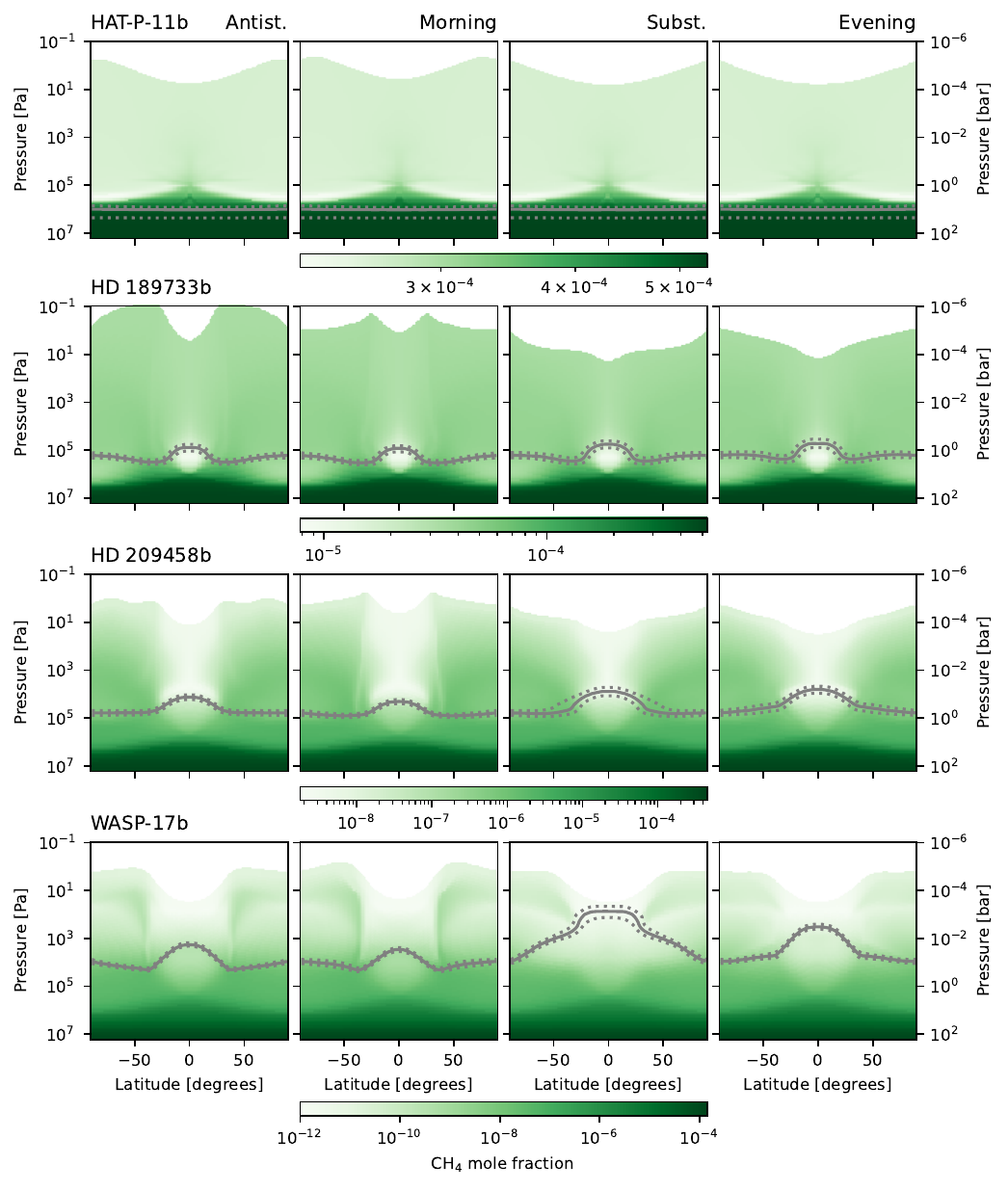}
    \caption{
    Pressure--latitude slices of the \ce{CH4} mole fraction
    at the antistellar, morning, substellar, and evening longitudes of the atmospheres of HAT-P-11b, HD~189733b, HD~209458b, and WASP-17b.
    The \ce{CH4} quench level is shown by a solid grey line, and the \ce{CH4} quench region by dotted grey lines.
    Colour bars differ between planets (one per row).
    }
    \label{fig:slices_pres_local_ch4_molefrac}
\end{figure*}

\begin{figure*}
    \centering
    \includegraphics[width=\textwidth]{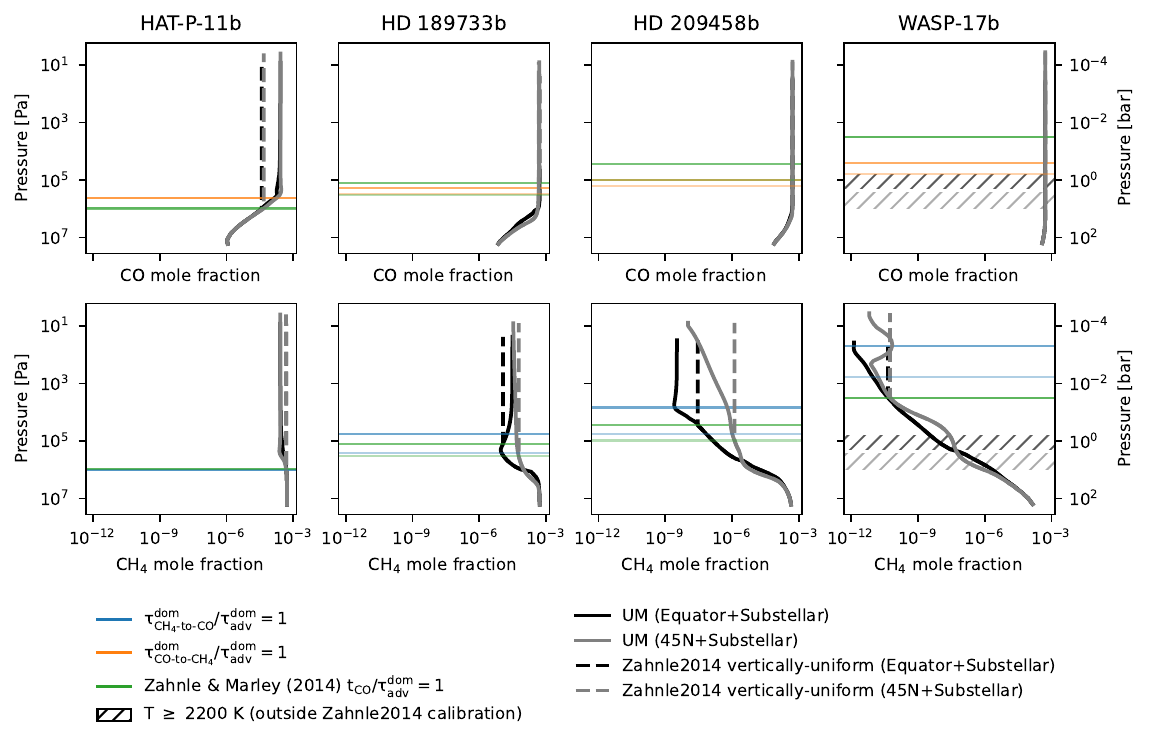}
    \caption{
    Vertical profiles of \ce{CO} (top) and \ce{CH4} (bottom) mole fraction on the substellar longitude at the equator (black) and \ang{45}N (grey), from the simulations (UM, solid) and from a quench treatment using $t_{\ce{CO}}$ (Zahnle2014, dashed), which holds abundances vertically uniform above the quench level.
    Horizontal lines mark the quench levels, where $\tau^{\mathrm{dom}}_{\ce{CH4-to-CO}}$ (blue, bottom row), $\tau^{\mathrm{dom}}_{\ce{CO-to-CH4}}$ (orange, top row) or $t_{\ce{CO}}$ (green) equals $\tau^{\mathrm{dom}}_{\mathrm{adv}}$; dark and light shades give the equator and \ang{45}N.
    Hatching marks $T \geq \SI{2200}{\kelvin}$, in the same shades.
    Corresponding pressure--latitude slices are given in Appendix~\ref{sec:appendix_zahnle_slices}.
    }
    \label{fig:um_vs_zahnle_vps}
\end{figure*}

Above the quench level \ce{CH4} abundance is governed by transport together with the slower but still active \ce{CH4}-to-\ce{CO} conversion chemistry (Section~\ref{sec:def_quenching}).
Transport carries \ce{CH4}-richer air into \ce{CH4}-poorer regions, while conversion pulls \ce{CH4} back towards its local equilibrium.
The resulting \ce{CH4} distribution depends on three factors: (a) the contrast \ce{CH4} quenching creates at the quench level (Section~\ref{sec:results_3d_mechanism}), (b) where transport carries the \ce{CH4}-richer air, and (c) whether conversion can remove that \ce{CH4} where the air arrives.
On HAT-P-11b quenching creates no contrast at the quench level, but transport lifts \ce{CH4}-richer air over the equator faster than conversion can remove it, so \ce{CH4} is enriched at the equator above the quench level.
On HD~189733b and the two hotter planets quenching depletes \ce{CH4} at the equator, but that depletion fares differently above the quench level.
On HD~189733b transport carries \ce{CH4}-richer air towards the equator \citep{Drummond18_3d,Drummond20_implications}, where conversion is too slow to remove it, and so fills in most of this depletion.
On HD~209458b and WASP-17b transport also carries \ce{CH4}-richer air towards the equator, but delivers it where conversion still removes it, so \ce{CH4} stays depleted at the equator above the quench level (Figures~\ref{fig:slices_pres_local_ch4_molefrac} and~\ref{fig:um_vs_zahnle_vps}; only the flanks of HD~209458b's jet gain \ce{CH4}).

In summary, our 3D GCM simulations show that transport together with the still-active conversion chemistry creates substantial spatial heterogeneity in \ce{CH4} above the quench level, even without photochemistry, which our network omits.

\subsection[Comparison with analytic conversion timescales]{Comparison with analytic conversion timescales}
\label{sec:results_zahnle}

We now ask how well a 1D quench treatment built on the timescale of \citet{Zahnle14_methane} reproduces our simulations, taking \ce{CO} and \ce{CH4} in turn.
A 1D quench treatment has three steps: first, it estimates a conversion timescale; second, it locates the quench level by comparing that conversion timescale with the transport timescale; and third, it holds the abundance vertically uniform above the quench level.
We assess each step, and find that the first two steps hold and the third fails.

\subsubsection[The Zahnle and Marley timescale]{The Zahnle \& Marley timescale}
\label{sec:results_zahnle_tco}

\citet{Zahnle14_methane} provide an analytic estimate of the \ce{CO}-to-\ce{CH4} conversion timescale, $t_{\ce{CO}}$, that is widely used to set quench levels in 1D models \citep[e.g.][]{Hu21_photochemistry,Mukherjee24_sonora,Mukherjee25_effects,Glein25_deciphering}.
$t_{\ce{CO}}$ is derived from an ensemble of 1D photochemical kinetics simulations spanning effective temperatures of \SIrange{500}{1100}{\kelvin}, gravities of \SIrange{e3}{e5}{\centi\metre\per\second\squared} and eddy diffusivities of \SIrange{e4}{e11}{\centi\metre\squared\per\second}.
The incident ultraviolet flux is set low enough that photolysis is negligible where quenching occurs.
In each simulation they located the quench point by comparing the reaction quotient $p_{\ce{CH4}}p_{\ce{H2O}}/p_{\ce{CO}}p_{\ce{H2}}^{3}$ against its equilibrium value (their Equations 5 and 6), taking the deepest level at which the two differ by more than ten per cent, then fitted an Arrhenius form to the quench points so obtained.
The fit therefore describes the network's emergent behaviour rather than the rate of any individual reaction, which is what makes it comparable with our full-pathway timescales (Section~\ref{sec:calc_tau_chem}).

Since $t_{\ce{CO}}$ describes the \ce{CO}-to-\ce{CH4} direction, we also compute $\tau^{\mathrm{dom}}_{\ce{CO-to-CH4}}$.
The \ce{CO}-to-\ce{CH4} pathway cannot be found on the graph of Section~\ref{sec:calc_tau_chem}, which keeps each reversible reaction only in its thermodynamically favoured direction.
That favoured direction is towards \ce{CO} at the quench level on all of our four planets, leaving no \ce{CO}-to-\ce{CH4} path through C--H--O species alone.\footnote{Paths routing carbon through N-bearing carriers do exist on the graph of Section~\ref{sec:calc_tau_chem}, but are some eight to nine orders of magnitude slower than the path found on the second graph and are not physically meaningful.}
We therefore build and search a second graph, in which the disfavoured direction of each reversible reaction is also present, at the rate given by its equilibrium constant.

We evaluate $t_{\ce{CO}}$ from Equations 12--14 of \citet{Zahnle14_methane}:
\begin{equation}
\label{eq:zahnle}
\begin{aligned}
    t_{\mathrm{q1}} &= 1.5\times10^{-6}\,p^{-1}m^{-0.7}\,e^{42000/T}, \\
    t_{\mathrm{q2}} &= 40\,p^{-2}\,e^{25000/T}, \\
    t_{\ce{CO}} &= \left(\frac{1}{t_{\mathrm{q1}}}+\frac{1}{t_{\mathrm{q2}}}\right)^{-1},
\end{aligned}
\end{equation}
\noindent where $p$ is in bar, $T$ in K, $t$ in seconds, and $m$ is metallicity relative to solar, so that $m=1$ for our simulations.
Across the \citet{Zahnle14_methane} ensemble of simulations, quench temperatures span roughly \SIrange{1000}{2200}{\kelvin} (their figure 5).
We exclude points warmer than \SI{2200}{\kelvin}, above which \citet{Zahnle14_methane} find no quenching anywhere in the ensemble.
Comparisons of this kind have been made in 1D by \citet{Tsai18_consistent} and \citet{Soni23_effect}.

\begin{figure}
    \centering
    \includegraphics[width=\columnwidth]{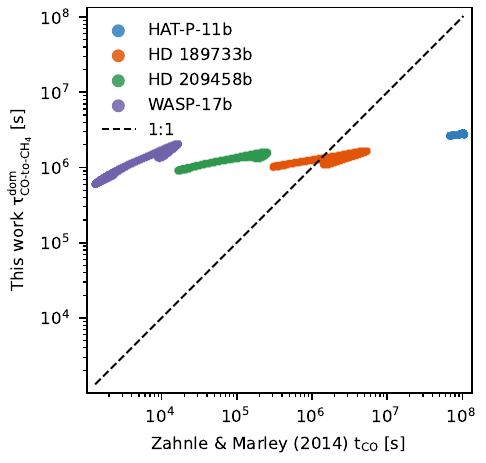}
    \caption{
    The dominant \ce{CO}-to-\ce{CH4} conversion timescale of this work, $\tau^{\mathrm{dom}}_{\ce{CO-to-CH4}}$, against the analytic estimate $t_{\ce{CO}}$ of \citet{Zahnle14_methane}, at every \ce{CO}-to-\ce{CH4} quench-level crossing on HAT-P-11b, HD~189733b, HD~209458b, and WASP-17b.
    This is the direction $t_{\ce{CO}}$ was fitted for.
    The dashed line marks equality; points below it are crossings at which $t_{\ce{CO}}$ is longer than $\tau^{\mathrm{dom}}_{\ce{CO-to-CH4}}$.
    }
    \label{fig:um_vs_zahnle_tau_co_ch4}
\end{figure}

\begin{figure}
    \centering
    \includegraphics[width=\columnwidth]{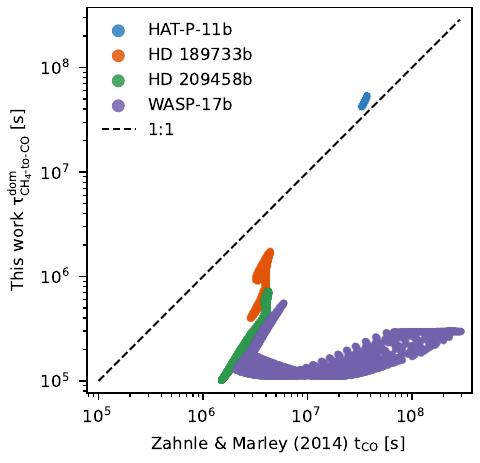}
    \caption{
    As Figure~\ref{fig:um_vs_zahnle_tau_co_ch4}, but for the \ce{CH4}-to-\ce{CO} direction: $\tau^{\mathrm{dom}}_{\ce{CH4-to-CO}}$ against $t_{\ce{CO}}$ at every \ce{CH4}-to-\ce{CO} quench-level crossing.
    Note the different axis ranges.
    }
    \label{fig:um_vs_zahnle_tau_ch4_co}
\end{figure}

$t_{\ce{CO}}$ is not a chemical rate.
It is an empirically fitted quantity which, when set equal to the vertical diffusion timescales of the \citet{Zahnle14_methane} ensemble of simulations, reproduces the quench points of that ensemble of simulations.
The $t_{\ce{CO}}$ fit was built for cloudless, self-luminous objects of near-solar composition, for which insolation is negligible and \ce{CH4} is thermodynamically favoured at observable depths.
\citet{Zahnle14_methane} are explicit that this restricts them to the \ce{CO}-to-\ce{CH4} direction, and that they do not treat the kinetic inhibition against oxidising \ce{CH4} to \ce{CO} --- the process that governs \ce{CH4} in strongly irradiated atmospheres, and the subject of this paper.
$t_{\ce{CO}}$ is nonetheless applied to temperate gas-rich planets \citep{Hu21_photochemistry}, across the Sonora Elf Owl grid \citep{Mukherjee24_sonora}, to irradiated atmospheres from gas giants to sub-Neptunes \citep{Mukherjee25_effects}, and at two hundred times solar metallicity \citep{Glein25_deciphering}.
We therefore compare $t_{\ce{CO}}$ against both directions (Figures~\ref{fig:um_vs_zahnle_tau_co_ch4} and~\ref{fig:um_vs_zahnle_tau_ch4_co}): $\tau^{\mathrm{dom}}_{\ce{CO-to-CH4}}$, which tests it on its own terms, and $\tau^{\mathrm{dom}}_{\ce{CH4-to-CO}}$, which is the direction with which this paper is concerned.

Each comparison is evaluated at its own quench level, defined as in Equation~(\ref{eq:quench_level}) with the corresponding conversion timescale.
For each vertical column we locate the two model levels between which the timescale ratio of Equation~(\ref{eq:quench_level}) crosses unity and interpolate log-linearly in pressure and timescale, and linearly in temperature, to the crossing.
Both timescales are evaluated at the same pressure and temperature, so the only difference between them is the conversion timescale itself.

\subsubsection[CO-to-CH4 conversion: the direction $t_{\ce{CO}}$ was fitted for]{CO-to-CH\textsubscript{4} conversion: the direction $t_{\ce{CO}}$ was fitted for}
\label{sec:results_zahnle_co}

For the \ce{CO}-to-\ce{CH4} conversion, the direction $t_{\ce{CO}}$ was fitted for, a quench treatment typically locates the \ce{CO} quench level within a factor of about four in pressure on every planet, and recovers the \ce{CO} abundance on every planet except HAT-P-11b.
Figure~\ref{fig:um_vs_zahnle_tau_co_ch4} compares $\tau^{\mathrm{dom}}_{\ce{CO-to-CH4}}$ with $t_{\ce{CO}}$ at every \ce{CO} quench-level crossing.
Table~\ref{tab:zahnle_co} follows the error through the three steps for each planet, and Figure~\ref{fig:um_vs_zahnle_vps} (top row) shows the resulting \ce{CO} profiles at the equator and \ang{45}N on the substellar longitude.
The median timescale offsets are large, up to a factor of 141 on WASP-17b, whose quench temperatures approach the upper limit of the fit.
These timescale offsets do not move the quench level as much, however, because the conversion timescale varies steeply with height: even on WASP-17b the median quench pressure shifts by a factor of about four.
Over WASP-17b's dayside the shift reaches a factor of about 20 (Figures~\ref{fig:um_vs_zahnle_vps} and~\ref{fig:um_vs_zahnle_slices_wasp17b}).

\begin{table}
    \centering
    \caption{Error through the three steps of a quench treatment of \ce{CO} based on $t_{\ce{CO}}$: the median offset of $t_{\ce{CO}}$ from $\tau^{\mathrm{dom}}_{\ce{CO-to-CH4}}$ at the quench-level crossings, the displacement of the \ce{CO} quench pressure, and the error in the vertically uniform \ce{CO} abundance at the top of the model domain, at two latitudes on the substellar longitude. All entries are factors. The displacement is the whole-grid median ratio of the $t_{\ce{CO}}$ quench pressure to the simulated one (below unity: shallower). The abundance error is the ratio of the vertically uniform abundance to the simulated one (below unity: underestimate).}
    \label{tab:zahnle_co}
    \begin{tabular}{lrrrr}
        \hline
        Planet & $t_{\ce{CO}}$ offset & $p_{\mathrm{qn}}$ error & \ce{CO} \ang{0}N & \ce{CO} \ang{45}N \\
        \hline
        HAT-P-11b & 32.092 & 2.240 & 0.145 & 0.176 \\
        HD~189733b & 1.571 & 1.121 & 1.039 & 0.948 \\
        HD~209458b & 8.031 & 0.607 & 1.000 & 0.998 \\
        WASP-17b & 141.375 & 0.230 & 0.998 & 0.998 \\
        \hline
    \end{tabular}
\end{table}

The \ce{CO} abundance error depends instead on whether \ce{CO} is the major or the minor carbon carrier.
On the three warmer planets \ce{CO} carries nearly all the carbon, so its profile is close to uniform above the quench level and holding it uniform introduces an error of about five per cent.
On HAT-P-11b \ce{CH4} is the major carrier at the quench level, and the vertically uniform \ce{CO} abundance is too low by a factor of six to seven.

\subsubsection[CH4-to-CO conversion: the direction irradiated atmospheres need]{CH\textsubscript{4}-to-CO conversion: the direction irradiated atmospheres need}
\label{sec:results_zahnle_ch4}

For the \ce{CH4}-to-\ce{CO} conversion, the direction irradiated atmospheres need, a quench treatment typically locates the \ce{CH4} quench level within a factor of two in pressure on every planet, but does not recover the \ce{CH4} abundance on any of them.
Figure~\ref{fig:um_vs_zahnle_tau_ch4_co} compares $\tau^{\mathrm{dom}}_{\ce{CH4-to-CO}}$ with $t_{\ce{CO}}$ at every \ce{CH4} quench-level crossing, Table~\ref{tab:zahnle_ch4} follows the error through the three steps for each planet, and Figure~\ref{fig:um_vs_zahnle_vps} (bottom row) shows the resulting \ce{CH4} profiles.
The median timescale offsets reach a factor of 15 on WASP-17b but move the median quench pressure by only about a factor of two, again because the conversion timescale varies steeply with height.
On HAT-P-11b the quench-pressure displacement is smaller than the width of the quench region.
Over WASP-17b's substellar tropics, however, the displacement reaches a factor of about 20 (Figure~\ref{fig:um_vs_zahnle_slices_wasp17b}), because the switch to pathway~(\ref{pathway4}) lifts the simulated quench level (Section~\ref{sec:results_3d_mechanism}) but $t_{\ce{CO}}$ does not follow that rise.

\begin{table}
    \centering
    \caption{As Table~\ref{tab:zahnle_co}, but for \ce{CH4}. The $t_{\ce{CO}}$ quench level is the same as in Table~\ref{tab:zahnle_co}: a single level serves both species.}
    \label{tab:zahnle_ch4}
    \begin{tabular}{lrrrr}
        \hline
        Planet & $t_{\ce{CO}}$ offset & $p_{\mathrm{qn}}$ error & \ce{CH4} \ang{0}N & \ce{CH4} \ang{45}N \\
        \hline
        HAT-P-11b & 1.375 & 0.934 & 1.875 & 1.845 \\
        HD~189733b & 3.436 & 1.372 & 0.381 & 1.773 \\
        HD~209458b & 7.225 & 1.685 & 8.773 & 121.317 \\
        WASP-17b & 15.243 & 2.032 & 33.249 & 8.628 \\
        \hline
    \end{tabular}
\end{table}

The \ce{CH4} abundance error is large because \ce{CH4} keeps changing above the quench level, where 3D transport and the slower but still active conversion chemistry continue to reshape that abundance (Section~\ref{sec:def_quenching}).
A quench treatment holds \ce{CH4} vertically uniform by construction, and so misestimates \ce{CH4} by up to two orders of magnitude on HD~209458b and WASP-17b, where \ce{CH4} is the minor carbon carrier.
On HAT-P-11b \ce{CH4} falls steadily between the quench level and the top of the domain, and the uniform profile overestimates it by a factor of about two.

On HD~189733b the \ce{CH4} abundance error also changes sign between latitudes: the uniform profile underestimates the \ce{CH4} abundance at the equator and overestimates it at \ang{45}N.
Above the quench level the \ce{CH4} abundance rises with height at the equator, where equatorward flow delivers \ce{CH4}-richer air, and falls with height at \ang{45}N (Section~\ref{sec:results_3d_above}).
A 1D treatment, with one profile to offer, cannot be right at both latitudes.

\subsection{Implications for chemical relaxation and 1D quench schemes}
\label{sec:results_zahnle_abundances}

To draw out what our results imply for chemical relaxation and 1D quench schemes, we start with the three steps of a 1D quench treatment (Section~\ref{sec:results_zahnle}).
On our four planets, the treatment typically locates the quench level but fails above it.
The first two steps hold up: median timescale offsets of up to a factor of 15 for \ce{CH4} and 141 for \ce{CO} move the median quench pressure by only about a factor of two and four respectively (Tables~\ref{tab:zahnle_ch4} and~\ref{tab:zahnle_co}), because the conversion timescale varies steeply with height.
The treatment fails at its third step, holding the abundance vertically uniform above the quench level, and that step is common to every quench scheme, not only to one built on the \citet{Zahnle14_methane} timescale.
The treatment misestimates the minor carbon carrier, whose abundance keeps changing above the quench level.
No choice of quench level can correct that error, because a single abundance per column cannot represent the structure that transport and conversion build above the quench level.
Such errors are not always visible from within a 1D framework: the Sonora Elf Owl grid \citep{Mukherjee24_sonora} quenched \ce{CO2} against the equilibrium \ce{CO} abundance rather than the \ce{CO} abundance above its own quench level, an error of one to two orders of magnitude in \ce{CO2} that came to light only against JWST spectra of cool brown dwarfs \citep{Wogan25_sonora}.

We now turn to both simplifications, chemical relaxation and 1D quench treatments, for which our pathway results (Section~\ref{sec:results_3d}) and quench-treatment results (Section~\ref{sec:results_zahnle}) matter differently.
In a GCM, a chemical relaxation scheme carries the abundance as a tracer, so the resolved flow sets its distribution above the quench level.
The scheme approximates only the chemistry, relaxing the abundance towards equilibrium on the timescale of a single conversion pathway, an approximation that holds where one pathway controls conversion at the quench level: on three of our four planets, but not on WASP-17b (Section~\ref{sec:results_3d_mechanism}).
A 1D quench treatment, by contrast, shares that single-timescale limitation and must also prescribe the abundance above the quench level, a step that fails on all four planets.
We cannot separate how much of this failure comes from assuming vertical uniformity and how much from the 1D geometry itself: a 1D kinetics model with a suitable $K_{zz}$ would retain some structure above the quench level, where conversion remains active.
The opposite-signed errors on HD~189733b (Section~\ref{sec:results_zahnle_ch4}), however, cannot be removed by any choice of $K_{zz}$, so at least part of the error is 3D.

\section{Discussion}
\label{sec:discussion}

\subsection{What quenching is}
\label{sec:disc_quenching}

Quenching is conventionally equated with a well-mixed profile above the quench level, an association that arose from the vertical-mixing geometry of its first application to a planetary atmosphere \citep{Prinn77_carbon}.
In that geometry the association holds: all air above the quench level arrives through the quench level and so carries the quench-level abundance, which makes the profile uniform.
In 3D, however, air above the quench level also arrives from other latitudes and longitudes, and conversion chemistry continues to act on it, so the profile need not be uniform.
On HD~189733b, for example, \ce{CH4} quenches, yet above the quench level the \ce{CH4} abundance rises with height at the equator and falls with height at \ang{45}N (Sections~\ref{sec:results_3d_above} and~\ref{sec:results_zahnle_ch4}).
We therefore define quenching by the transfer of control from thermochemistry to transport (Section~\ref{sec:def_quenching}), not by the profile it leaves behind.
This definition separates what quenching sets from what happens above the quench level.
Quenching sets the abundance at the quench level, and transport together with the still-active conversion chemistry shapes its distribution above.

The inherited picture also takes the quench level itself to be spatially uniform.
In our simulations the quench level is 3D: on HD~189733b and HD~209458b it sits higher at the equator than at higher latitudes, and on WASP-17b it varies throughout the atmosphere (Section~\ref{sec:results_3d_mechanism}).
\citet{Zamyatina24_quenchingdriven} inferred similar behaviour on WASP-96b, where increasing metallicity also altered the quench level structure.

\subsection{Uncertainties in 3D modelling of quenching}
\label{sec:disc_uncertainties}

\textbf{Single-timescale schemes.} Our results support previous studies highlighting the importance of solving thermochemical kinetics in 3D models of hot-Jupiter atmospheres \citep{Drummond20_implications,Lee23_minichemical,Zamyatina23_observability,Zamyatina24_quenchingdriven}.
Our WASP-17b simulation, in particular, shows that a switch in the dominant \ce{CH4}-to-\ce{CO} conversion pathway reshapes the quench level and the resulting \ce{CH4} distribution.
Schemes built on a single conversion timescale, whether chemical relaxation in GCMs or quench treatments in 1D models, would miss such a switch, which could bias \ce{CH4} and \ce{CO} predictions and the inferred \ce{C}/\ce{O} ratio, metallicity and intrinsic temperature.
On our three cooler planets, however, where a single pathway controls \ce{CH4} quenching, relaxation schemes remain adequate.

\textbf{Analytic timescales.} For 1D quench treatments, the analytic timescale $t_{\ce{CO}}$ of \citet{Zahnle14_methane} is off by a median factor of up to 15 on our planets (Table~\ref{tab:zahnle_ch4}).
That timescale error is consistent with \citet{Tsai18_consistent}, who found that a simple Arrhenius-like fit fails to reproduce the \ce{CH4}-to-\ce{CO} conversion timescale across the conditions relevant for hot Jupiters.
Our pathways suggest why: a pathway rich in termolecular steps, such as (\ref{pathway1}), and one carrying only the final dissociation, such as (\ref{pathway4}), cannot share a common effective activation energy and pressure exponent.
Even so, because the conversion timescale varies steeply with height, $t_{\ce{CO}}$ typically locates the \ce{CH4} quench level within a factor of two in pressure (Section~\ref{sec:results_zahnle}).
The exception is WASP-17b's substellar tropics, where $t_{\ce{CO}}$ does not follow the switch to pathway~(\ref{pathway4}) and the quench level is displaced by a factor of about 20 (Section~\ref{sec:results_zahnle_ch4}).

\textbf{Chemical network.} Our \ce{CH4}-to-\ce{CO} conversion pathways and timescales depend on the adopted network \citep{Tsai18_consistent}.
The other network now coupled to the \texttt{UM}, the \texttt{minichem} network of \citeauthor{Tsai22_minichemical}~(\citeyear{Tsai22_minichemical}; benchmarked against ours by \citealt{Christie26_benchmarking}), cannot test our pathways by construction.
\texttt{minichem}'s net reactions carry the rate of the rate-limiting step of the pathway they replace, so the pathway itself is not resolved, whereas $\tau^{\mathrm{dom}}_{\ce{CH4-to-CO}}$ uses the full-pathway rate (Section~\ref{sec:calc_tau_chem}).
Testing our pathways will therefore require coupling other networks that resolve individual reactions to the same GCM.
New networks of this kind have been developed \citep{Veillet24_extensively,Veillet26_development}, but must first be reduced for use in hot-Jupiter GCMs.

\textbf{\ce{CH3OH} thermochemistry.} \ce{CH3OH} is the key intermediate in all our pathways except (\ref{pathway4}) and (\ref{pathway7}), among them (\ref{pathway1}), which operates at the \ce{CH4} quench level on every planet.
Although \ce{CH3OH} chemistry is relatively well known, remaining uncertainties have been shown to affect \ce{CH4} quenching behaviour in 1D photochemical kinetics simulations of the atmosphere of HD~189733b \citep{Venot12_chemical,Moses14_chemical}.
\citet{Moses14_chemical} traced those uncertainties to the reaction \ce{CH3OH + H -> CH3 + H2O}, whose adopted rate coefficient \citep{Hidaka89_thermal} is probably too fast.
This reaction does not appear in any of our dominant pathways, which instead pass through \ce{CH3OH + H -> CH3O + H2} or \ce{CH3OH + H -> CH2OH + H2}, but it competes with both for \ce{CH3OH}.
The \citet{Venot19_reduced} network we use retains that fast rate coefficient from \citet{Venot12_chemical}, whose \ce{CH3OH} chemistry \citet{Venot20_new} have since revised.
\citet{Venot20_new} found little change for hot Jupiters, but \citet{Tsai21_comparative} found that the revised chemistry raises the \ce{CH4} abundance at the quench level, so \ce{CH4} above the quench level may be biased low in our simulations.

\textbf{Transport.} \citet{Cooper06_dynamics} distinguished the transport that sets the quench level from the transport that shapes the abundance above it: in their simulations of HD~209458b, vertical advection sets the quench level, and horizontal mixing then homogenises \ce{CO}.
Other studies find other directions setting the quench level: zonal advection \citep[for WASP-43b,][]{Mendonca18_threedimensional}, or horizontal and vertical advection together \citep[for HD~209458b and HD~189733b,][]{Drummond18_observable,Drummond18_3d}.
We find zonal advection setting the \ce{CH4} quench level on HAT-P-11b, and zonal and vertical advection comparably fast on HD~189733b, HD~209458b and WASP-17b.
Our advection timescales rest on chosen length scales (Section~\ref{sec:calc_tau_adv}) and are dayside averages, so unlike the conversion timescale they do not vary in 3D; forthcoming work will use a locally defined advection timescale.

\textbf{Deep thermal structure.} Our simulations use a fixed intrinsic temperature of \SI{100}{\kelvin} at the bottom boundary, and their deep atmosphere is unlikely to have reached thermal convergence within the simulation runtime \citep{Sainsbury-Martinez19_idealised}.
True deep temperatures are therefore likely higher than simulated, most consequentially on HAT-P-11b and HD~189733b, where \ce{CH4} quenches deepest.
Warmer deep temperatures would shorten the conversion timescale on these two planets, raising the quench level and lowering the quench-level \ce{CH4} abundance (Section~\ref{sec:results_3d_mechanism}), and could also change the dominant pathway.
We defer a systematic exploration of this sensitivity to future work.

\textbf{Unresolved convection.} Because we apply no dry convective adjustment scheme, tracer mixing by unresolved convective motions is not represented.
That omission is minor in the radiation-dominated region probed by observed spectra, where the resolved flow dominates vertical mixing.
Unresolved convection may matter, however, where \ce{CH4} quenches deep, on HAT-P-11b and HD~189733b.
Extra vertical mixing would push the quench level deeper, towards higher equilibrium \ce{CH4} abundance, opposite to the effect of warmer deep temperatures.
HD~189733b is the more exposed, because its vertical advection already rivals zonal advection at the quench level.
On HAT-P-11b vertical advection lags zonal advection by more than an order of magnitude, so unresolved convection would have to close that gap to change the quench level.

\textbf{Photochemistry.} Our network contains no photolysis, so photochemical destruction of \ce{CH4} on the irradiated dayside \citep{Moses14_chemical,Tsai21_comparative,Baeyens22_grid} is not represented.
On the three cooler planets \ce{CH4} quenches too deep for photolysis to reach, so photolysis would deplete \ce{CH4} only above the quench level, on the dayside.
On WASP-17b, however, the substellar quench level rises to \SI{2e2}{\pascal} (Section~\ref{sec:results_equator}), at the longitude receiving the most ultraviolet flux.
At such low pressure, photolysis could compete with the interconversion chemistry and alter the quench condition itself.

\subsection{Quenching as a diagnostic of atmospheric processes}

Quenching provides a useful but non-trivial probe of deep atmospheric processes \citep{Fortney20_beyond,Sing24_warm}.
Because quenching often occurs well below the observable photosphere, \ce{CH4} above the quench level can carry information about conditions at depth.
On HAT-P-11b and HD~189733b that information survives: above the quench level \ce{CH4} stays within a factor of about three of its quench-level abundance (Figure~\ref{fig:um_vs_zahnle_vps}).
On HD~209458b and WASP-17b, however, conversion keeps removing \ce{CH4} above the quench level, by up to two orders of magnitude (Sections~\ref{sec:results_3d_above} and~\ref{sec:results_zahnle_ch4}), so the observable \ce{CH4} no longer traces conditions at depth.
Our 3D simulations suggest that 1D modelling of cooler atmospheres, as in \citet{Fortney20_beyond} and \citet{Sing24_warm}, works for HAT-P-11b: the quench level is flat, so observable \ce{CH4} barely varies with latitude or longitude.
For HD~189733b, however, 1D modelling works for the wrong reason: observable \ce{CH4} is nearly as uniform, but only because 3D transport fills in the equatorial depletion at the quench level (Section~\ref{sec:results_3d_above}).

Poorly constrained metallicity further complicates \ce{CH4} as a diagnostic of conditions at depth.
On WASP-96b, a planet not analysed here, higher metallicity warms the atmosphere at the quench depth while leaving the jet nearly unchanged.
The faster conversion raises the quench level, turning a weak equatorial \ce{CH4} depletion into a strong one \citep{Zamyatina24_quenchingdriven}.

Quenching and the transport above the quench level also provide the baseline against which photochemistry and aerosol formation should be identified.
Once transport takes control, abundances no longer follow local equilibrium, yet remain structured in 3D (Section~\ref{sec:results_3d_above}).
Quenching weakens the limb contrasts that thermochemical equilibrium would impose, but does not erase them \citep{Zamyatina23_observability,Zamyatina24_quenchingdriven}.
In our simulations the contrasts that survive are modest beside the limb asymmetries and day--night contrasts predicted for photochemical products \citep{Tsai23_day,Tsai24_global,Baeyens24_photodissociation}, and beside the transit-depth asymmetries attributed to aerosols \citep{Espinoza24_inhomogeneous,Fu25_overcast,Mak25_impact}.
Those quenching contrasts still have to be modelled before an observed asymmetry can be attributed to either process.

\section{Conclusions}
\label{sec:conclusions}

We explored the spatial variability of \ce{CH4}-to-\ce{CO} conversion pathways in aerosol-free, solar-composition Met Office \texttt{Unified Model} GCM simulations coupling hydrodynamics, radiative transfer and thermochemical kinetics (excluding photolysis) for the atmospheres of HAT-P-11b, HD~189733b, HD~209458b and WASP-17b.
We used these simulations to test the two simplifications adopted in place of 3D kinetics: chemical relaxation in GCMs and 1D quench treatments.

Our conclusions are as follows:
\begin{itemize}
\item[(1)] We identify ten dominant \ce{CH4}-to-\ce{CO} conversion pathways across our simulations, varying with depth, longitude and planet.
\item[(2)] Despite this diversity, a single pathway controls \ce{CH4} quenching on HAT-P-11b, HD~189733b, and HD~209458b, indicating that quenching in these atmospheres can be characterised by a single conversion timescale.
Chemical relaxation schemes should therefore be adequate for these three atmospheres, at least within the constraints of our GCM.
\item[(3)] On WASP-17b, by contrast, two pathways compete at the \ce{CH4} quench level, and the switch between them lifts the quench level over the substellar tropics, deepening the equatorial \ce{CH4} depletion.
Schemes built on a single conversion timescale are inadequate for WASP-17b.
\item[(4)] At the \ce{CH4} quench level, conversion competes with zonal advection on HAT-P-11b, and with zonal and vertical advection on the other three planets; meridional advection is slower on all four.
\item[(5)] The \ce{CH4} quench level varies in 3D because two effects shorten the conversion timescale, and so raise the quench level: (a) a hotter atmosphere and (b) a faster pathway.
Because \ce{CH4} is less abundant at equilibrium higher up, a higher quench level means a lower quench-level \ce{CH4} abundance.
That rise of the quench level is the mechanism behind the equatorial \ce{CH4} depletion at the quench level on HD~189733b, HD~209458b and WASP-17b.
\item[(6)] Above the quench level, transport together with the still-active conversion chemistry reshapes \ce{CH4}.
The resulting distribution depends on three factors: (a) the contrast \ce{CH4} quenching creates at the quench level, (b) where transport carries \ce{CH4}-richer air, and (c) whether conversion can remove that \ce{CH4} where the air arrives.
On HD~189733b conversion is too slow to remove the arriving \ce{CH4}, so transport fills in most of the equatorial depletion; on HD~209458b and WASP-17b conversion removes that \ce{CH4}, so the depletion survives.
On HAT-P-11b quenching creates no contrast, yet transport lifts \ce{CH4}-richer air over the equator faster than conversion can remove it, enriching the equator.
Structure above the quench level therefore needs neither photochemistry nor a varying quench level.
\item[(7)] A 1D quench treatment built on the \citet{Zahnle14_methane} timescale typically locates the quench level within a factor of about two in pressure for \ce{CH4} and four for \ce{CO} on every planet, despite median timescale offsets of up to 15 and 141 (Tables~\ref{tab:zahnle_ch4} and~\ref{tab:zahnle_co}).
The treatment does not recover the abundance above the quench level: held vertically uniform, the minor carbon carrier is wrong by up to two orders of magnitude, and on HD~189733b in opposite directions at the equator and \ang{45}N, which no choice of $K_{zz}$ can correct.
That failure lies in holding the abundance vertically uniform, a step common to every quench scheme, not in the \citet{Zahnle14_methane} timescale.
\item[(8)] Observable \ce{CH4} traces conditions at depth on HAT-P-11b and HD~189733b, where \ce{CH4} above the quench level stays within a factor of about three of its quench-level abundance.
On HD~209458b and WASP-17b, however, conversion keeps removing \ce{CH4} above the quench level, by up to two orders of magnitude, so observable \ce{CH4} no longer traces conditions at depth.
1D modelling works for HAT-P-11b, whose quench level is flat, but works for HD~189733b only for the wrong reason, because 3D transport fills in the equatorial depletion.
\item[(9)] More broadly, our results caution against the common picture of abundances above the quench level as static, ``frozen-in'' values: quenching sets an abundance at the quench level, but does not fix it thereafter.
\end{itemize}

Together, these results show that solving chemical kinetics in 3D matters both for predicting disequilibrium abundances in hot-Jupiter atmospheres and for interpreting \ce{CH4} signatures in their spectra.
Quenching and the transport above the quench level also provide the baseline against which photochemistry and aerosol formation should be identified in those spectra.

\section*{Acknowledgments}

MZ and NJM were supported by a UKRI Future Leaders Fellowship MR/T040866/1 and its extension MR/Z000122/1, and by the Leverhulme Trust through a research project grant RPG-2020-82.
ÉH, RV, and NJM acknowledge support from a Science and Technology Facilities Council Small Award ST/Y00261X/1.
DAC received funding from the Max Planck Society.
HB was supported by a Science and Technology Facilities Council Studentship ST/Y509383/1.

This research was performed using the DiRAC Data Intensive service at Leicester, operated by the University of Leicester IT Services, which forms part of the STFC DiRAC HPC Facility (\url{www.dirac.ac.uk}). The equipment was funded by BEIS capital funding via STFC capital grants ST/K000373/1 and ST/R002363/1 and STFC DiRAC Operations grant ST/R001014/1. DiRAC is part of the National e-Infrastructure.

Material produced using Met Office Software.
Data analysis made use of the following \texttt{Python} packages: \texttt{aeolus} \citep{aeolus}, \texttt{ipython} \citep{ipython}, \texttt{iris} \citep{iris}, \texttt{jupyter} \citep{jupyternotebook}, \texttt{matplotlib} \citep{matplotlib}, \texttt{networkx} \citep{networkx}, \texttt{numpy} \citep{numpy}, and \texttt{quenchpath} \citep{quenchpath}.

\section*{Data Availability}

The research data supporting this publication are openly available from the Open Research Exeter (ORE) online repository at \url{https://doi.org/10.24378/exe.4304}. For the purpose of open access, the authors have applied a Creative Commons Attribution (CC BY) licence to any Author Accepted Manuscript version arising.

\bibliographystyle{mnras}
\bibliography{references}

\appendix

\section[Pressure--latitude structure behind the quench level analysis]{Pressure--latitude structure behind the quench level analysis}
\label{sec:appendix_quench_slices}

\begin{figure*}
    \centering
    \includegraphics[width=\textwidth]{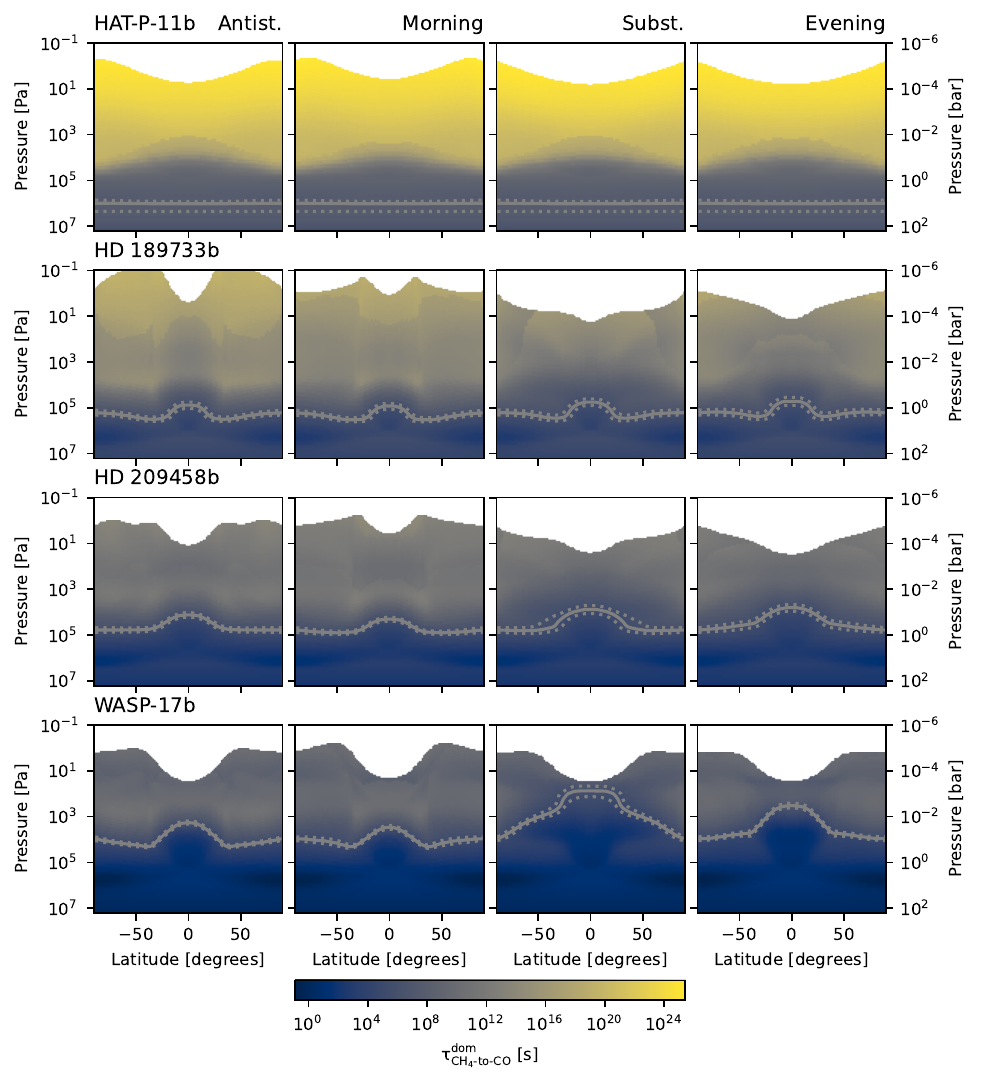}
    \caption{
    Same as in Figure~\ref{fig:slices_pres_local_pathways} but for the dominant \ce{CH4}-to-\ce{CO} conversion timescale.
    }
    \label{fig:slices_pres_local_tau_ch4_net}
\end{figure*}

\begin{figure*}
    \centering
    \includegraphics[width=\textwidth]{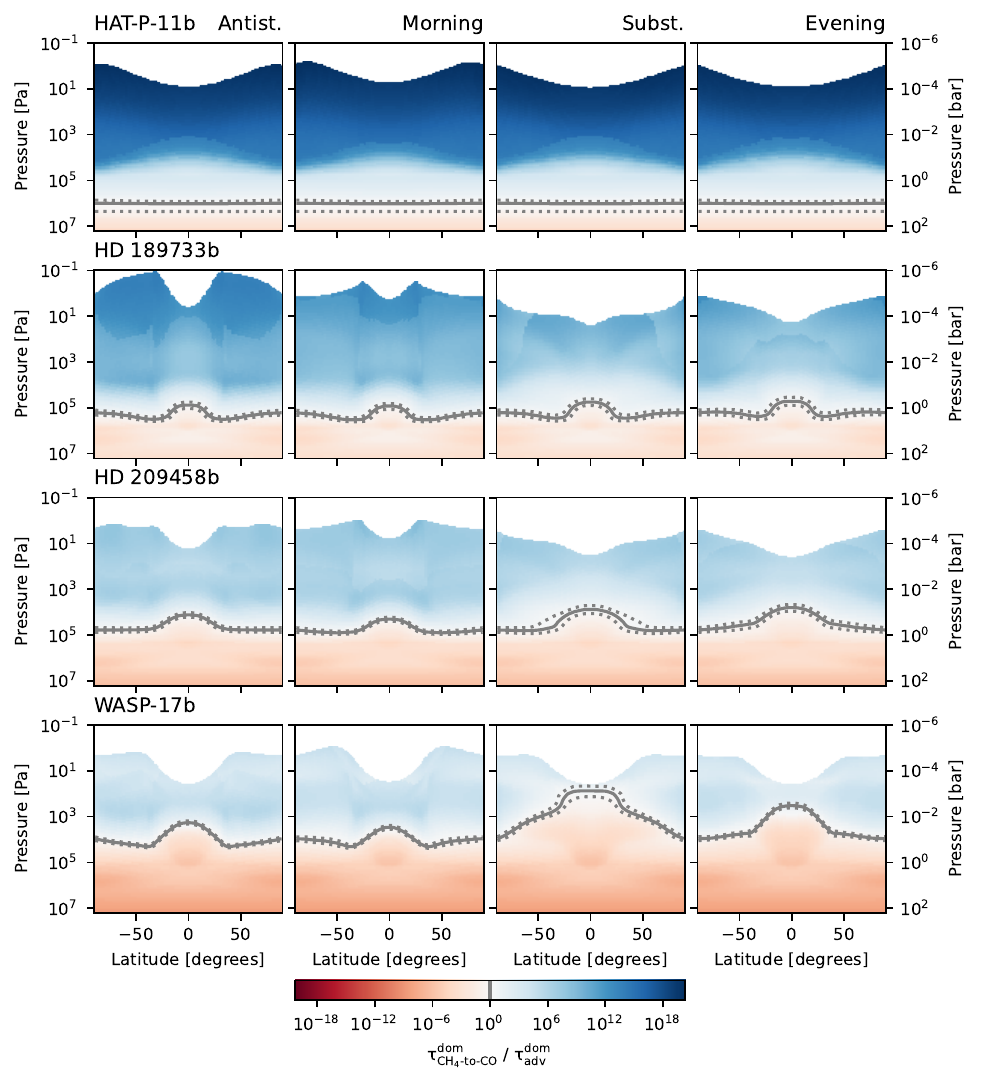}
    \caption{
    Same as in Figure~\ref{fig:slices_pres_local_pathways} but for the ratio of the dominant \ce{CH4}-to-\ce{CO} conversion timescale to the dominant advection timescale (Equation~\ref{eq:quench_level}).
    }
    \label{fig:slices_pres_local_tau_ch4_net_over_tau_ls_dom}
\end{figure*}

\begin{figure*}
    \centering
    \includegraphics[width=\textwidth]{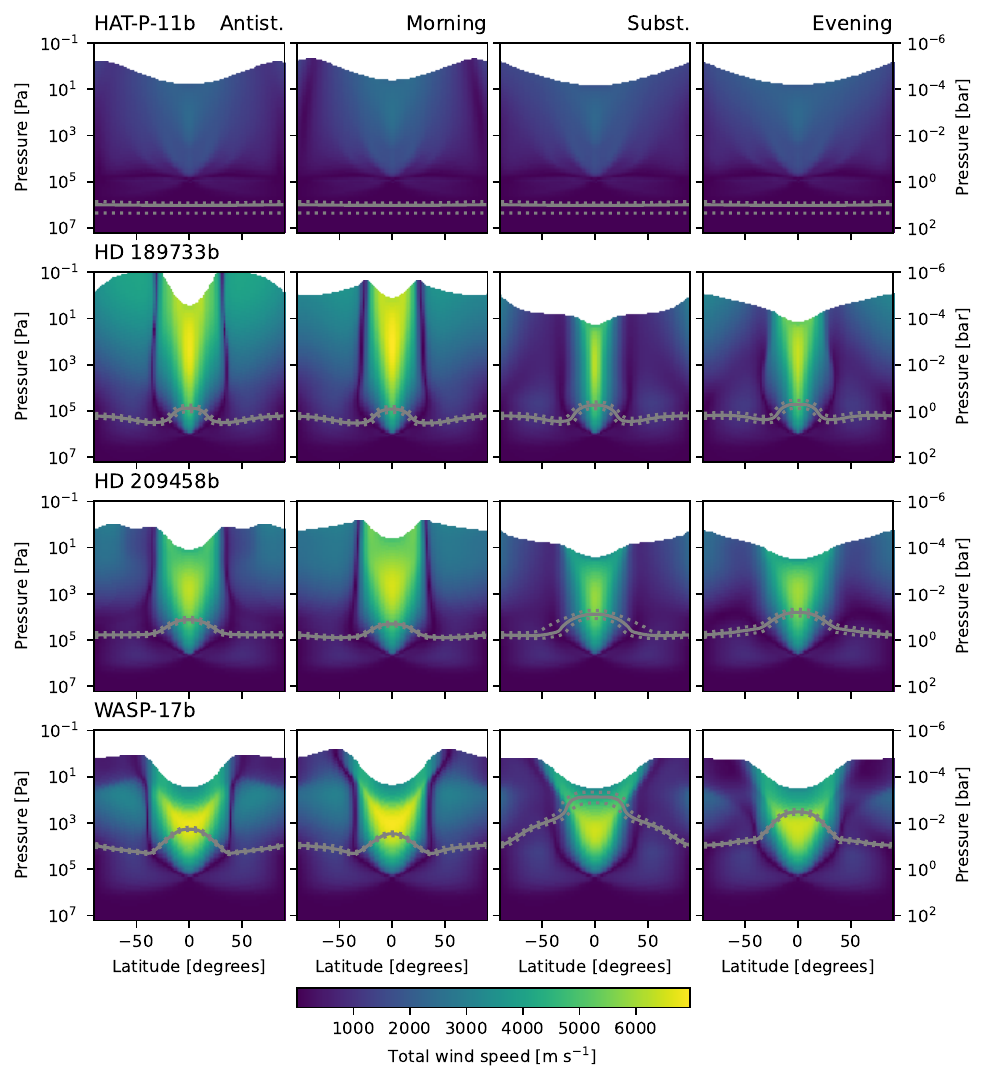}
    \caption{Same as in Figure~\ref{fig:slices_pres_local_ch4_molefrac} but for the total wind speed, with a single colour bar for all planets.}
    \label{fig:slices_pres_local_wind}
\end{figure*}

\begin{figure*}
    \centering
    \includegraphics[width=\textwidth]{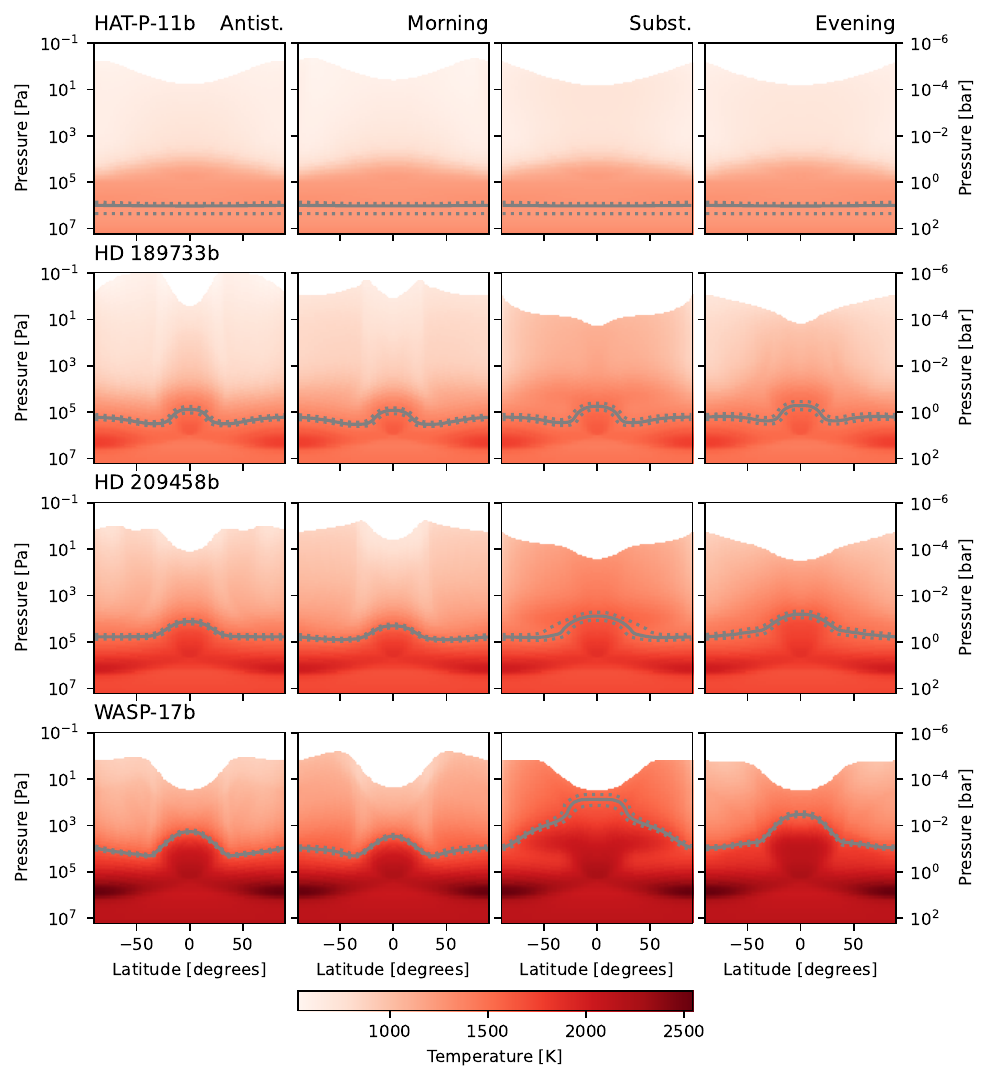}
    \caption{Same as in Figure~\ref{fig:slices_pres_local_ch4_molefrac} but for the gas temperature, with a single colour bar for all planets.}
    \label{fig:slices_pres_local_temp}
\end{figure*}

This appendix collects the pressure--latitude figures underlying Section~\ref{sec:results_3d}.
Figures~\ref{fig:slices_pres_local_tau_ch4_net} and~\ref{fig:slices_pres_local_tau_ch4_net_over_tau_ls_dom} show the dominant \ce{CH4}-to-\ce{CO} conversion timescale and its ratio to the dominant advection timescale; the \ce{CH4} quench level lies where that ratio equals unity (Equation~\ref{eq:quench_level}).
Figures~\ref{fig:slices_pres_local_wind} and~\ref{fig:slices_pres_local_temp} show the total wind speed and the gas temperature, which set the advection and the conversion timescale respectively.

\section[Pressure--latitude comparison with Zahnle \& Marley (2014)]{Pressure--latitude comparison with Zahnle \& Marley (2014)}
\label{sec:appendix_zahnle_slices}

\begin{figure*}
    \centering
    \includegraphics[width=\textwidth,trim=0 6 0 6,clip]{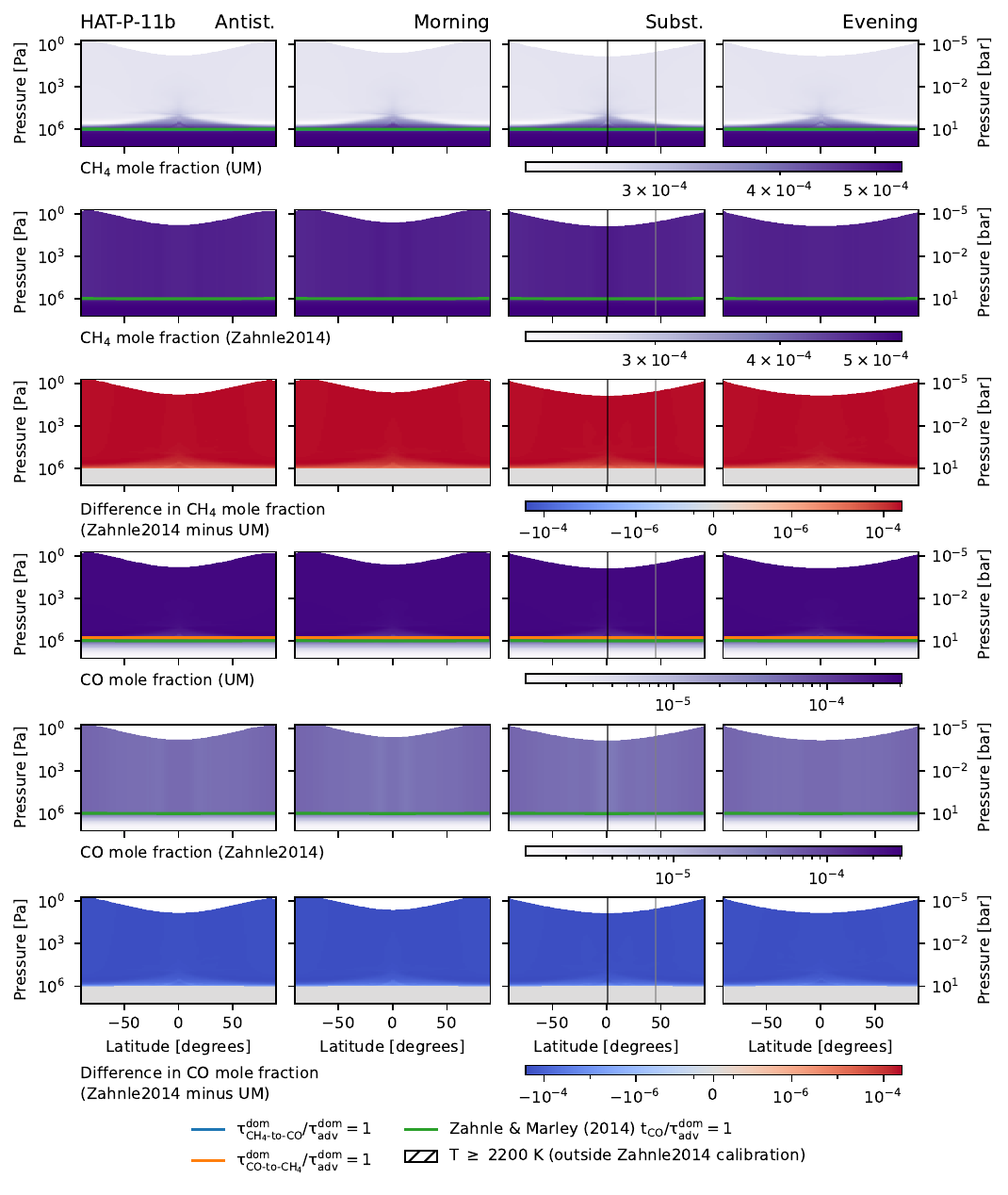}
    \caption{
    Pressure--latitude slices behind Figure~\ref{fig:um_vs_zahnle_vps}, for HAT-P-11b.
    For each species, the top row shows the mole fraction from the simulation (UM), the middle row that from the quench treatment using $t_{\ce{CO}}$ (Zahnle2014), and the bottom row their difference.
    The quench treatment holds the abundance vertically uniform above the quench level.
    \ce{CH4} occupies the top three rows and \ce{CO} the bottom three; columns show the antistellar, morning, substellar, and evening longitudes.
    Lines on the simulation and quench-treatment rows mark the quench levels, where $\tau^{\mathrm{dom}}_{\ce{CH4-to-CO}}$ (blue, \ce{CH4} rows), $\tau^{\mathrm{dom}}_{\ce{CO-to-CH4}}$ (orange, \ce{CO} rows) or $t_{\ce{CO}}$ (green) equals $\tau^{\mathrm{dom}}_{\mathrm{adv}}$.
    Vertical lines on the substellar column mark the two latitudes of Figure~\ref{fig:um_vs_zahnle_vps}.
    Hatching marks $T \geq \SI{2200}{\kelvin}$, outside the calibration of \citet{Zahnle14_methane}.
    }
    \label{fig:um_vs_zahnle_slices_hatp11b}
\end{figure*}

\begin{figure*}
    \centering
    \includegraphics[width=\textwidth,trim=0 6 0 6,clip]{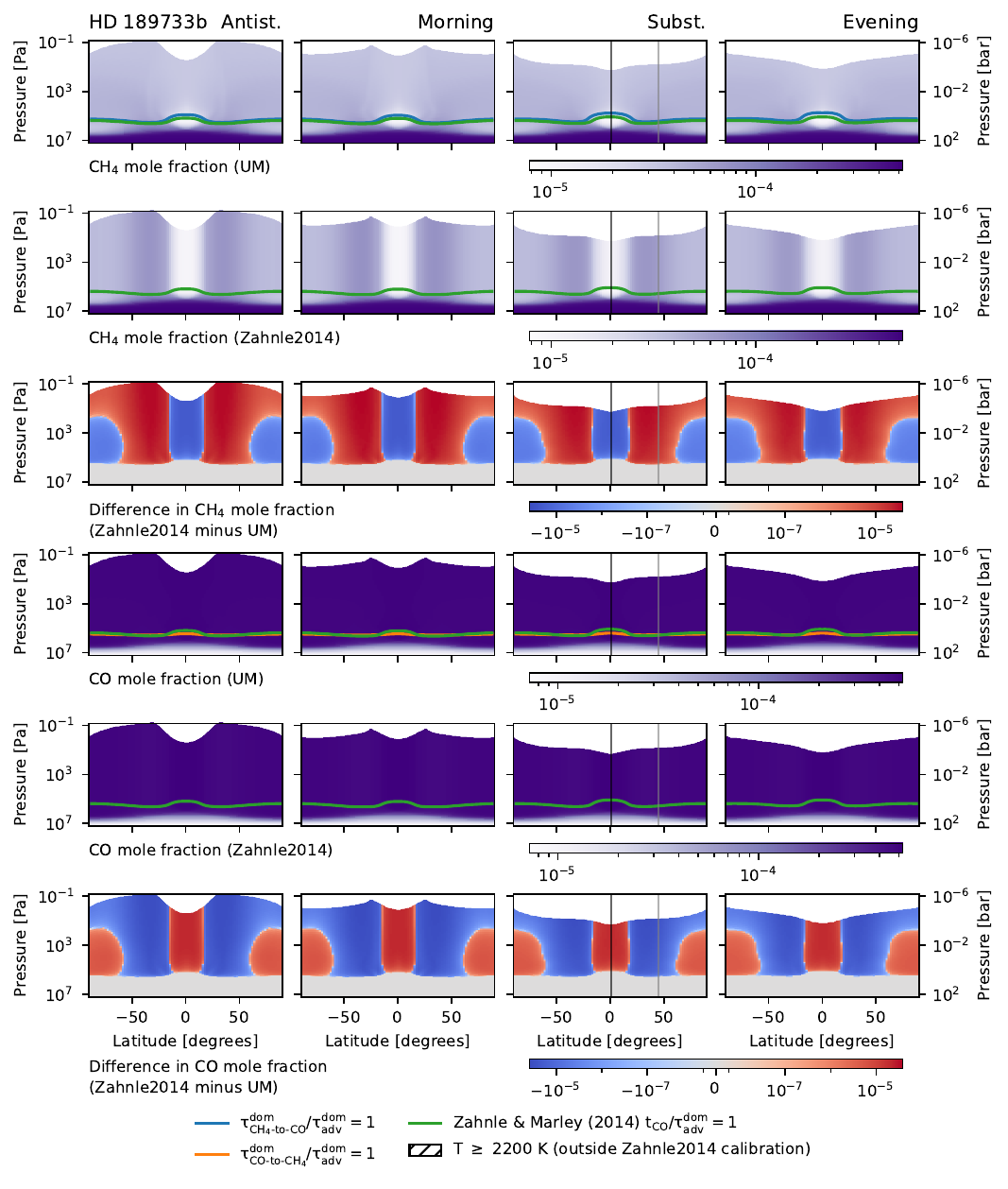}
    \caption{Same as in Figure~\ref{fig:um_vs_zahnle_slices_hatp11b} but for HD~189733b.}
    \label{fig:um_vs_zahnle_slices_hd189733b}
\end{figure*}

\begin{figure*}
    \centering
    \includegraphics[width=\textwidth,trim=0 6 0 6,clip]{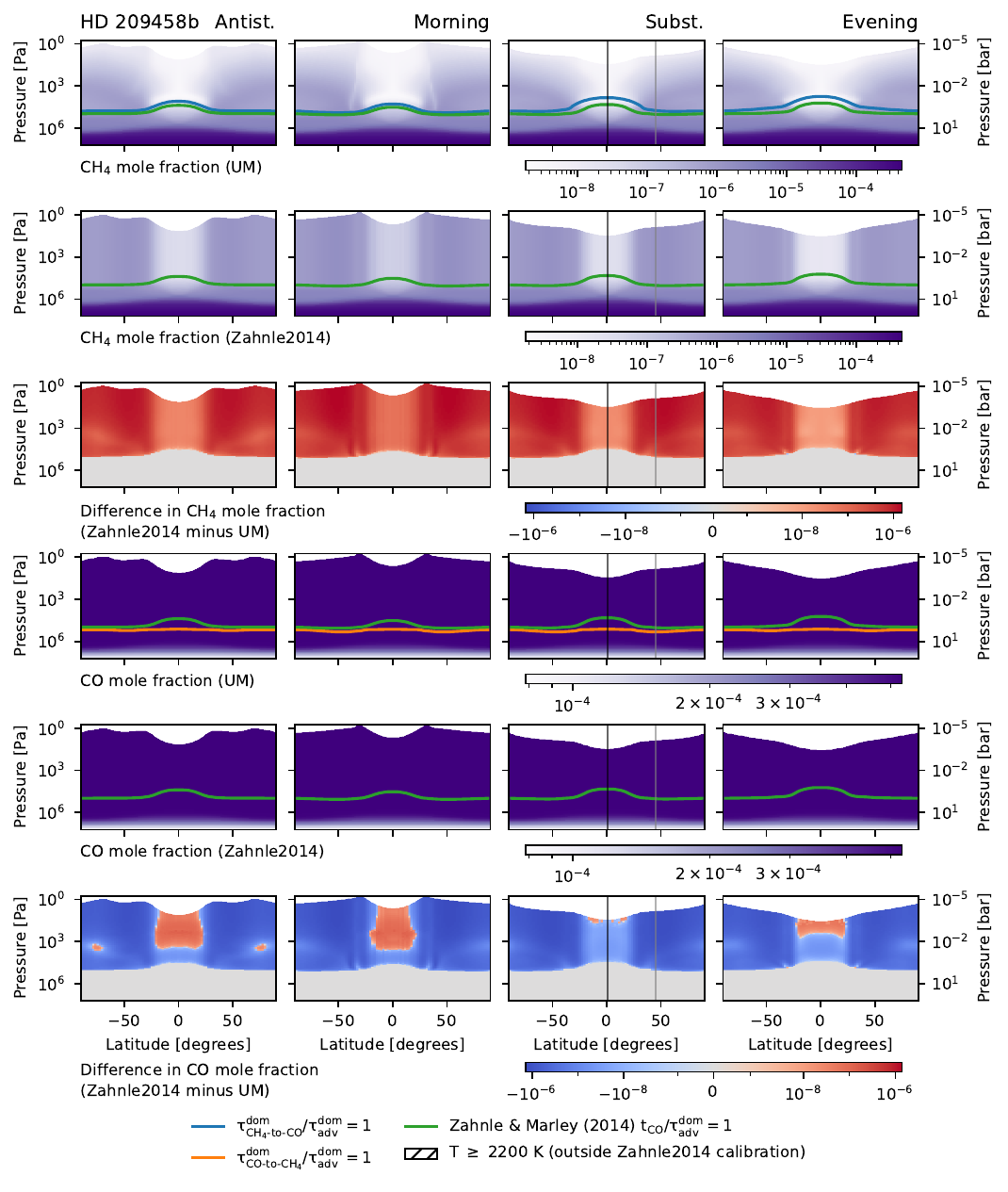}
    \caption{Same as in Figure~\ref{fig:um_vs_zahnle_slices_hatp11b} but for HD~209458b.}
    \label{fig:um_vs_zahnle_slices_hd209458b}
\end{figure*}

\begin{figure*}
    \centering
    \includegraphics[width=\textwidth,trim=0 6 0 6,clip]{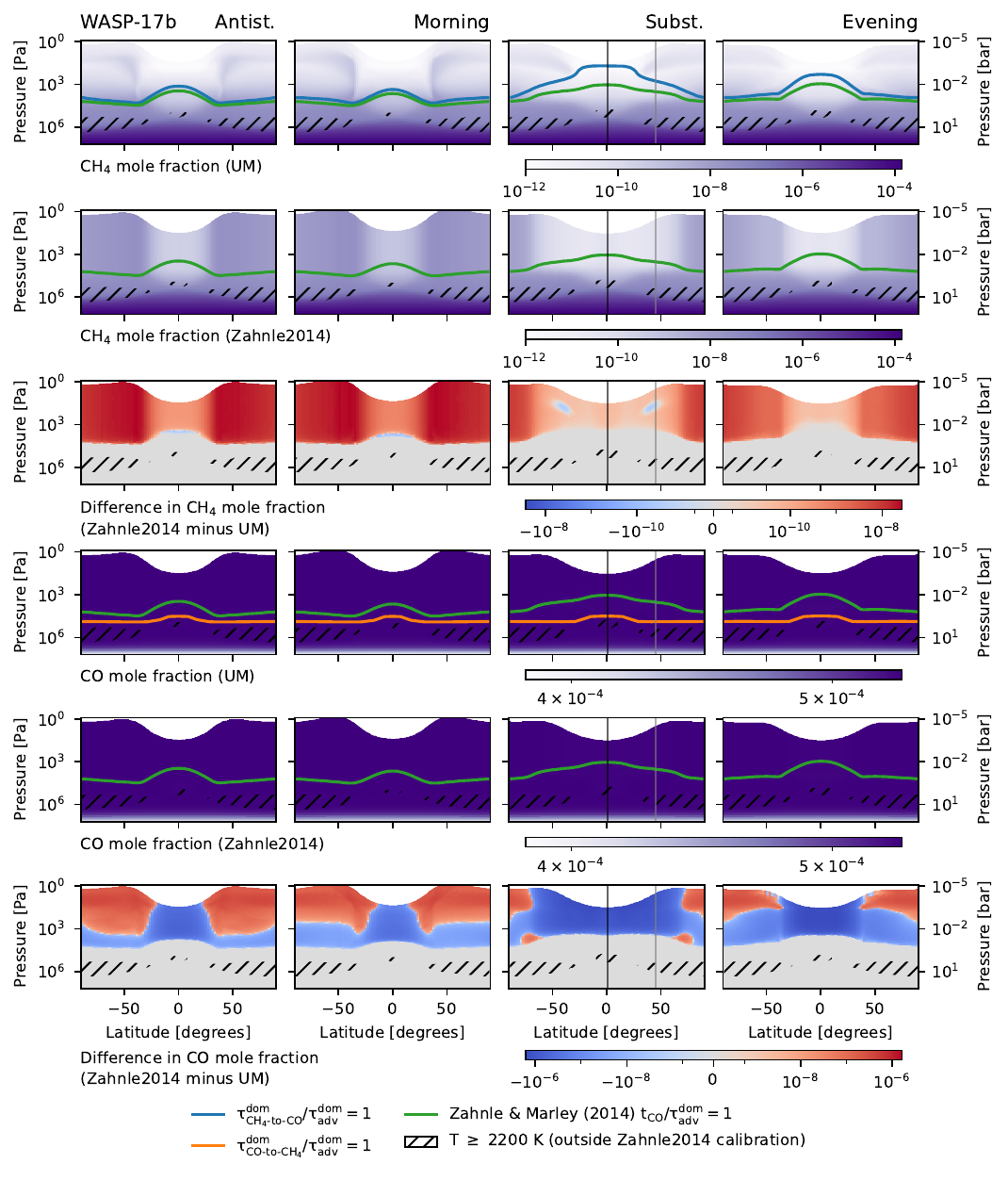}
    \caption{Same as in Figure~\ref{fig:um_vs_zahnle_slices_hatp11b} but for WASP-17b.}
    \label{fig:um_vs_zahnle_slices_wasp17b}
\end{figure*}

Figure~\ref{fig:um_vs_zahnle_vps} compares the simulated (UM) and quench-treatment (Zahnle2014) profiles at two latitudes on a single longitude.
Figures~\ref{fig:um_vs_zahnle_slices_hatp11b}--\ref{fig:um_vs_zahnle_slices_wasp17b} show the same comparison as pressure--latitude slices at all four longitudes, one figure per planet, so that the spatial dependence of the error is visible rather than sampled at two points.
On HAT-P-11b (Figure~\ref{fig:um_vs_zahnle_slices_hatp11b}) the difference rows (Zahnle2014 minus UM) are of one sign: the vertically uniform profile overestimates \ce{CH4} and underestimates \ce{CO}.
On HD~189733b (Figure~\ref{fig:um_vs_zahnle_slices_hd189733b}) the sign of the \ce{CH4} error instead changes with latitude (Table~\ref{tab:zahnle_ch4}), which no single 1D profile can reproduce.
Figures~\ref{fig:um_vs_zahnle_slices_hd209458b} and~\ref{fig:um_vs_zahnle_slices_wasp17b} show the corresponding structure for HD~209458b and WASP-17b.

\bsp
\label{lastpage}
\end{document}